%% file: main.tex
\pdfoutput=1
\documentclass[preprint,12pt]{elsarticle}

\usepackage{hyperref}
\usepackage{amsmath}
\usepackage{amssymb}
\usepackage{bm}
\usepackage{graphicx}
\usepackage{booktabs}
\usepackage{multirow}
\usepackage{subcaption}
\usepackage{algorithm}
\usepackage{algpseudocode}
\usepackage{cleveref}
\usepackage[T1]{fontenc}
\usepackage{lmodern}
\usepackage[utf8]{inputenc}
\usepackage{siunitx}
\usepackage{microtype}
\usepackage{float}
\usepackage[section]{placeins}
\hypersetup{hidelinks,pageanchor=false}
\newcommand{\xvec}{\mathbf{x}}
\newcommand{\vvec}{\mathbf{v}}
\newcommand{\avec}{\mathbf{a}}

\newcommand{\Emesh}{\mathcal{E}_{\mathrm{mesh}}}

\newcommand{\Vfree}{V_{\mathrm{free}}}

\begin{document}

\sloppy

\begin{frontmatter}

\title{Crash Assessment via Mesh-Based Graph Neural Networks and Physics-Aware Attention}

\author[seat]{Gabriel Curtosi\corref{cor1}}
\ead{gabriel.curtosi@seat.es}
\author[idiada]{Carlos Manuel Ruiz Ruiz}
\author[seat]{Fabiola Cavaliere}
\author[seat]{Xabier Larráyoz Izcara}

\address[seat]{SEAT S.A., Barcelona, Spain}
\address[idiada]{IDIADA Automotive Technology S.A., Santa Oliva, Spain}

\cortext[cor1]{Corresponding author}

\begin{abstract}
Full-vehicle crash simulations are computationally expensive, limiting their use in iterative design exploration. This work investigates learned hybrid surrogate models (\textit{MeshTransolver}, \textit{MeshGeoTransolver}, and \textit{MeshGeoFLARE}) for predicting time-resolved structural deformation fields in an industrial lateral pole-impact benchmark. We evaluate whether neural surrogates can reproduce full-field crash kinematics with sufficient accuracy, spatial regularity, and structural plausibility for engineering interpretation. The proposed architectures combine local mesh message passing, geometry-aware global attention, and sparse contact-aware correction for autoregressive crash rollout.

We compare mesh-based graph neural networks, attention-based geometric models, and hybrid architectures under a common training and hyperparameter configuration. The hybrid models capture both short-range structural interactions and long-range deformation patterns, while a sparse contact-aware variant assesses the effect of dynamic proximity interactions during rollout.

On a 25-sample full-vehicle test set, the best hybrid model achieves a temporal mean root-mean-square error of 3.20 mm. While geometry-aware attention baselines are quantitatively competitive, qualitative side-view inspection shows they can introduce local spatial noise and deformation irregularities that complicate structural interpretation. In contrast, hybrid mesh-attention models provide the best balance between scalar accuracy, survival-space consistency, and physically interpretable displacement fields.

These results suggest that crash surrogate assessment should combine global error metrics with downstream safety-relevant quantities and qualitative field inspection. The proposed methodology enables fast full-field predictions while preserving essential structural information for industrial crash-engineering analysis.
\end{abstract}


\end{frontmatter}

\input{01_introduction}

\input{02_problem_dataset}

\input{03_related_work}

\input{04_method}

\input{05_experiments}

\input{06_results}

\input{07_discussion}

\input{08_conclusion}

\section*{Acknowledgements}
The authors gratefully acknowledge SEAT S.A. for providing access to the full-vehicle structural FE model and simulation infrastructure, as well as for the use of its high-performance computing resources.

Part of this work was developed using the NVIDIA PhysicsNeMo open-source framework~\citep{physicsnemo2025}. The authors also thank Willian Patrick Da Silva Rodrigues for his valuable support in preparing the vehicle-crash dataset.

\section*{Funding}
This research did not receive any specific grant from funding agencies in the public, commercial, or not-for-profit sectors. Structural FE models and HPC resources were provided internally by SEAT S.A.

\section*{Data availability}
The full structural finite-element models and raw simulation datasets used in this work are proprietary and confidential and cannot be publicly shared.

\section*{Declaration of generative AI and AI-assisted technologies in the manuscript preparation process}
During the preparation of this work, the authors used ChatGPT 5.4 and Claude Opus 4.6 to support language editing, technical proofreading, and readability improvement of the manuscript. After using these tools, the authors reviewed and edited all generated content as needed and take full responsibility for the final content of the publication.

\section*{Declaration of competing interests}
The authors declare that they have no known competing financial interests or
personal relationships that could have appeared to influence the work reported
in this paper.

\bibliographystyle{elsarticle-num}
\bibliography{refs}

\end{document}

%% file: 01_introduction.tex
\section{Introduction}
\label{sec:introduction}

Vehicle structural safety is governed by stringent international standards, which require the preservation of occupant-compartment integrity under multiple crash configurations. In industrial practice, this requirement is addressed through extensive finite-element (FE) simulation campaigns using explicit solvers such as PAM-CRASH or LS-DYNA, complemented by physical testing for certification and correlation~\citep{euroncap2023,nhtsa2022}. Although simulation drives the development process and decision-making, each full-vehicle crash analysis remains expensive, which limits the number of design variants that can be explored within realistic project timelines.

Accurate prediction of vehicle structural response under impact is critical for safety assessment. Crash response is strongly nonlinear and typically involves large deformation, evolving contact, and long-range load transfer across the structure. These characteristics make crash simulation attractive for neural surrogates, but also challenging: the model must preserve local structural detail while remaining stable under closed-loop temporal prediction. In the present work, we focus on lateral pole impact as a representative full-vehicle load case that concentrates these difficulties in a severe and industrially relevant setting.

Neural surrogate models offer the potential for orders-of-magnitude acceleration, reducing evaluation time from hours to seconds during inference. Two main families have emerged for mesh-based physical simulation. Graph neural networks (GNNs) explicitly exploit mesh connectivity and provide strong local inductive bias~\citep{pfaff2021learning,sanchez2020learning}, but their locality can limit efficient propagation of long-range structural dependencies. Transformer-style scientific surrogates compress the state into latent tokens and model global interactions efficiently~\citep{wu2024transolver,adams2025geotransolver}, but often smooth out local deformation patterns when used without a mesh-based backbone. This trade-off motivates hybrid local-global architectures.

\begin{figure}[!t]
    \centering
    \includegraphics[width=\linewidth,height=0.85\textheight,keepaspectratio]{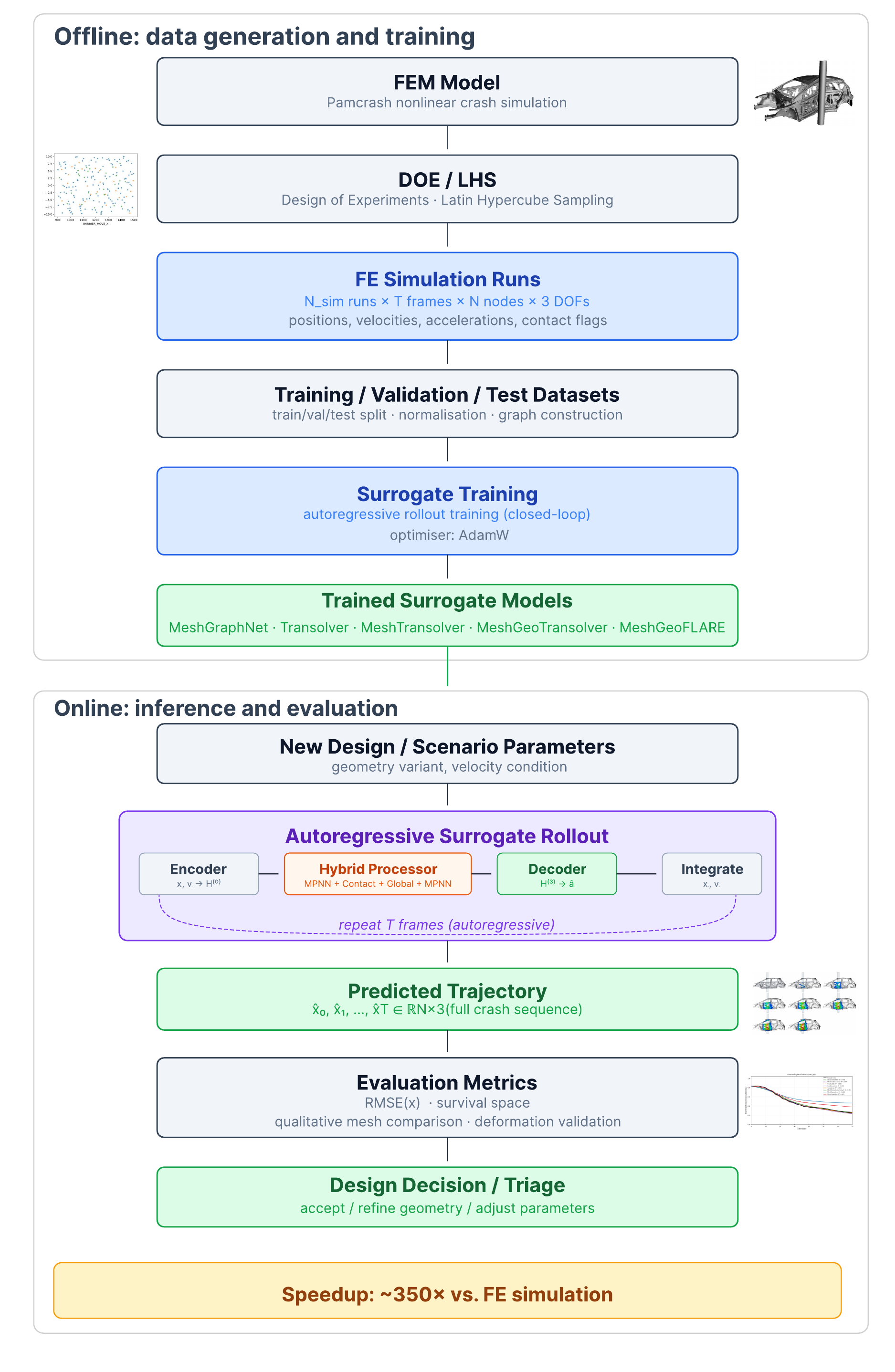}
    \caption{End-to-end pipeline of the proposed framework. High-fidelity finite-element simulations are replaced by autoregressive neural surrogate inference, reducing evaluation time from hours to seconds while preserving safety-critical evaluation.}
    \label{fig:pipeline}
\end{figure}

The proposed workflow in this paper is shown in Figure~\ref{fig:pipeline}. A parameterised full-vehicle finite-element model is sampled with Latin hypercube sampling to generate a design-space dataset, which is then simulated with a high-fidelity crash solver. The resulting trajectories are used to train neural surrogates that operate under fully autoregressive rollout at inference time. The deployment objective is not only to reproduce nodal kinematics, but also to preserve downstream safety-relevant quantities such as occupant survival space.

A further challenge arises from contact. In crash simulation, important interactions are not fully described by the static structural mesh, because proximity relations evolve during impact. In a fully autoregressive surrogate, these dynamic interactions must be represented in a way that is informative enough to improve accuracy, yet sufficiently controlled to avoid destabilising the rollout. This is one of the main design problems addressed in the present work.

This paper makes the following contributions:
\begin{enumerate}
  \item We formulate full-vehicle lateral pole-impact prediction as a fully autoregressive mesh-based surrogate problem and evaluate it on an industrial DOE dataset with statistical split validation.
  \item We introduce \textit{MeshTransolver} and its contact-aware variants, which combine local message passing with global physics-aware attention and explicit contact handling for crash simulation.
  \item We introduce an explicit generic sparse contact block for the autoregressive hybrid architectures, injecting dynamic contact information before the global processor while preserving mesh-only refinement afterwards.
  \item We extend the hybrid formulation with geometry-aware global processors and contact-aware injection, yielding \textit{MeshGeoTransolver} and \textit{MeshGeoFLARE}, where FLARE provides a more efficient global-attention layer.
  \item We evaluate all models using both standard RMSE global field errors and \textbf{occupant survival space} as a downstream safety-relevant industrial metric.
\end{enumerate}

The remainder of the paper is organised as follows.  Section~\ref{sec:problem_dataset} formulates the prediction problem, autoregressive rollout, and downstream safety metric. Section~\ref{sec:related_work} reviews the relevant background on local, global, and geometry-aware surrogate models. Section~\ref{sec:method} presents the proposed hybrid architectures and the sparse contact block. Section~\ref{sec:experiments} describes the full-vehicle dataset, design space, split validation, and training protocol. Section~\ref{sec:results} reports the full-vehicle results, including contact effects and occupant survival-space evaluation. Section~\ref{sec:discussion} discusses the main architectural lessons and remaining limitations. Section~\ref{sec:conclusion} concludes the paper.

%% file: 02_problem_dataset.tex
\section{Problem Formulation}
\label{sec:problem_dataset}

\subsection{Mesh-based crash forecasting}
\label{sec:formulation}

Let $\mathcal{G}=(\mathcal{V},\Emesh)$ denote the structural mesh graph, where $\mathcal{V}$ is the set of mesh nodes and $\Emesh$ is the edge set derived from FE connectivity. At each timestep $t\in\{0,1,\ldots,T\}$, the state of the structure is described by nodal positions $\xvec_t\in\mathbb{R}^{N\times 3}$ and velocities $\vvec_t\in\mathbb{R}^{N\times 3}$, with velocities estimated by finite differences,
\begin{equation}
  \vvec_t = \frac{\xvec_t-\xvec_{t-1}}{\Delta t}.
  \label{eq:velocity}
\end{equation}
The surrogate model $f_\theta$ predicts a normalised acceleration field from the current state,
\begin{equation}
  \bar{\avec}_t = f_\theta\!\bigl(\xvec_t,\,\vvec_t,\,\boldsymbol{\phi},\,\Emesh,\,\mathcal{C}(t)\bigr),
  \label{eq:surrogate}
\end{equation}
where $\boldsymbol{\phi}$ denotes static node features and $\mathcal{C}(t)$ denotes optional time-varying contact information.

\subsection{Autoregressive rollout}
\label{sec:rollout_problem}

All models are trained and evaluated under closed-loop autoregressive rollout. After denormalisation, the predicted acceleration is integrated with a forward Euler update,
\begin{equation}
  \vvec_{t+1}=\vvec_t+\Delta t\,\avec_t,\qquad
  \xvec_{t+1}=\xvec_t+\Delta t\,\vvec_{t+1},
  \label{eq:euler}
\end{equation}
and the predicted state is fed back as input for the next timestep. This matches the intended deployment setting, where the surrogate replaces the FE solver over the whole crash horizon rather than performing isolated one-step prediction.

The rollout process used during training and inference is illustrated later in Figure~\ref{fig:rollout}. When contact-aware variants (detailed in Section~\ref{sec:method}) are used, the sparse contact representation is recomputed from the predicted configuration during the rollout itself.

\subsection{Training objective}
\label{sec:loss}

The training objective is the mean-squared position error over deformable nodes and over the full rollout horizon,
\begin{equation}
  \mathcal{L}_{\mathrm{pos}} = \frac{1}{|\Vfree|\,T}\sum_{t=1}^{T}\sum_{n\in\Vfree}
  \bigl\|\hat{\xvec}_{n,t}-\xvec^*_{n,t}\bigr\|_2^2,
  \label{eq:loss}
\end{equation}
where $\xvec^*_{n,t}$ denotes the FE reference trajectory. Rigid nodes do not move during the simulation and are therefore excluded from both loss computation and state update, so $\Vfree\subset\mathcal{V}$ contains only deformable nodes.

\subsection{Downstream safety metric: occupant survival space}
\label{sec:survival_metric}

In addition to nodal field errors, we evaluate a downstream safety-relevant metric derived from the predicted structural response. The occupant survival space is defined as the Euclidean distance between two reference nodes located on the inner B-pillar and the central tunnel, which serves as a proxy for occupant compartment intrusion. Let
\begin{equation}
  d_t = \bigl\|\mathbf{x}_{A,t} - \mathbf{x}_{B,t}\bigr\|_2,
\end{equation}
denote this distance at time $t$, where points $A$ and $B$ correspond to the inner B-pillar and central tunnel, respectively.

The survival-space error $e^{\mathrm{surv}}_t$ is then defined as
\begin{equation}
  e^{\mathrm{surv}}_t = d^{\mathrm{pred}}_t - d^{\mathrm{GT}}_t,
  \label{eq:survival}
\end{equation}
where $d^{\mathrm{pred}}_t$ and $d^{\mathrm{GT}}_t$ denote the predicted and finite-element ground-truth (GT) distances, respectively.

This quantity is \emph{not} used as a training target, but as a downstream evaluation metric chosen for its industrial relevance. The sign of $e^{\mathrm{surv}}_t$ is critical: positive values ($e^{\mathrm{surv}}_t > 0$) indicate that the surrogate predicts a larger residual survival distance than the FE reference, which may be non-conservative if interpreted as available occupant space. Conversely, negative values ($e^{\mathrm{surv}}_t < 0$) correspond to a conservative overestimation of the true intrusion. Correct interpretation of this sign is essential for safety-critical design screening.

%% file: 03_related_work.tex
\section{Background and Related Work}
\label{sec:related_work}

This section summarises the main methodological ingredients that motivate the proposed architecture family. The objective is to clarify the complementary strengths and limitations of local mesh-based propagation, global token-based attention, geometry-aware attention mechanisms, and hybrid local-global processors for industrial crash-surrogate modelling.

\subsection{Graph neural networks for physical simulation}
\label{sec:rw_gnn}

Graph neural networks (GNNs) have become a standard choice for learning on discretised physical systems because they operate directly on irregular domains and inherit a strong neighbourhood-based inductive bias. Sanchez-Gonzalez et al.~\citep{sanchez2020learning} introduced \emph{Graph Network-based Simulators} (GNS), showing that encode--process--decode GNNs with autoregressive rollout can learn high-quality dynamics for fluids, rigid bodies, and deformable materials from trajectory data. Pfaff et al.~\citep{pfaff2021learning} extended this idea to mesh-based simulation with MeshGraphNet (MGN), which is the direct local-graph predecessor of the architectures considered in the present work.

MGN uses mesh connectivity as the graph structure and applies message-passing neural network (MPNN) blocks to propagate information along mesh edges. Each processor block performs an edge update followed by a node update:
\begin{align}
  \mathbf{m}_{ij}^{(\ell+1)} &= f_e\!\bigl(
    \mathbf{h}_i^{(\ell)}, \mathbf{h}_j^{(\ell)}, \mathbf{e}_{ij}^{(\ell)}\bigr),
    \label{eq:mgn_edge_rw} \\
  \mathbf{h}_i^{(\ell+1)} &= f_n\!\bigl(
    \mathbf{h}_i^{(\ell)},\; \textstyle\sum_{j \in \mathcal{N}(i)}
    \mathbf{m}_{ij}^{(\ell+1)}\bigr),
    \label{eq:mgn_node_rw}
\end{align}
where $\mathbf{h}_i^{(\ell)}$ denotes the latent state of node $i$ at processor layer $\ell$, $\mathbf{e}_{ij}^{(\ell)}$ denotes the latent edge feature associated with edge $(i,j)$, $\mathcal{N}(i)$ is the neighbourhood of node $i$ in the mesh graph, and $f_e$ and $f_n$ are learned update functions, typically implemented as MLPs with residual connections and normalisation.

For crash simulation, this locality is attractive because neighbouring FE elements directly constrain how deformation, force transfer, and structural interactions propagate through the body. As a result, the local mesh connectivity provides a strong structural prior that message-passing models can exploit effectively. This makes GNNs well suited to capturing short-range structural interactions, boundary effects, and local geometric detail.

At the same time, purely local propagation becomes less efficient when relevant information must travel across large graph diameters or when contact activates interactions between regions that were previously disconnected. In such settings, long-range dependencies must be recovered through multiple successive local updates, which can make learning more difficult. More generally, with $L$ MPNN layers, each node can only aggregate information from its $L$-hop neighbourhood. For large industrial FE meshes with mesh diameter $D \gg L$, global structural modes cannot propagate within a single forward pass without stacking many layers, which increases both computational cost and optimisation difficulty.

In summary, mesh-based MPNNs provide a strong local inductive bias and respect the structural adjacency encoded by the FE mesh, but their communication range is fundamentally limited by the number of message-passing steps.

\subsection{Global token-based attention on irregular domains}
\label{sec:rw_transolver}

A complementary line of work replaces purely local mesh propagation with global token-based attention on irregular domains. \textbf{Transolver}~\citep{wu2024transolver} introduces \emph{Physics Attention} as a scalable mechanism specifically designed for PDE learning on unstructured geometries. The central idea is to compress nodal states into a smaller set of learnable physics-aware tokens, perform global interaction in token space, and then map the updated representation back to the nodes.

Given a latent nodal matrix $\mathbf{H} \in \mathbb{R}^{N \times d}$ and a learned key matrix $\mathbf{K} \in \mathbb{R}^{M \times d}$, the slice weights and token matrix can be written as
\begin{align}
  \mathbf{W} &= \mathrm{Softmax}(\mathbf{H}\mathbf{K}^{\top}) \in \mathbb{R}^{N \times M},
  \label{eq:physattn_slice} \\
  \mathbf{T} &= \mathbf{W}^{\top}\mathbf{H} \in \mathbb{R}^{M \times d},
  \label{eq:physattn_token}
\end{align}
where $N$ is the number of nodes, $M \ll N$ is the number of learned physics tokens, and $d$ is the latent feature dimension. Self-attention is then applied on the $M$ tokens to produce $\widetilde{\mathbf{T}}$, and the updated nodal representation is recovered through a deslicing step:
\begin{equation}
  \widetilde{\mathbf{H}} = \mathbf{W}\,\widetilde{\mathbf{T}} \in \mathbb{R}^{N \times d}.
  \label{eq:physattn_deslice}
\end{equation}

By applying global interaction in token space rather than directly over all nodes, the method reduces the cost of dense communication while preserving access to long-range dependencies. This is attractive for large irregular domains, where many relevant couplings extend well beyond the receptive field of a shallow MPNN stack.

However, Transolver operates directly on nodal point sets rather than explicitly exploiting mesh connectivity. This removes the local structural prior provided by the FE mesh and can make it harder to preserve sharp neighbourhood-level detail at contact boundaries, interfaces, or highly localised deformation patterns.

\textbf{Transolver$^{++}$}~\citep{luo2025transolverpp} further extends this family with more distinguishable latent states and a refined Physics-Attention parameterisation for large-scale geometries. In the present work, this sharper token-based formulation is used as the basis for all Transolver-based hybrid variants.

Thus, token-based physics attention offers a complementary route to large-scale communication on irregular domains, but without the explicit mesh-local inductive bias of graph-based processors.

\subsection{Geometry-aware and efficient attention variants}
\label{sec:rw_geo}

Recent extensions strengthen token-based attention in two directions that are particularly relevant to industrial CAE problems: improved sensitivity to geometry and operating context, and improved computational efficiency at large scale.

\textbf{GeoTransolver}~\citep{adams2025geotransolver} augments the Transolver backbone with \emph{Geometry-Aware Latent Embedding} (GALE). Conceptually, GALE conditions token formation and latent interaction on learned geometry-aware context, so that the global processor becomes more sensitive to spatial configuration, loading location, and other global or boundary-condition information. This is especially relevant when the design space includes geometric variation and when the surrogate must remain responsive to changes in shape or structural layout.

Efficient attention variants such as \textbf{FLARE}~\citep{puri2025flare} target a different limitation: the memory and compute burden of dense global interaction. Rather than forming full dense attention patterns, FLARE uses a more efficient routed or factorised latent interaction that preserves global communication while reducing the cost of attention on large irregular domains.

Accordingly, \textbf{GeoTransolver} corresponds to a geometry-aware token-based processor, whereas \textbf{GeoFLARE} in the present work preserves the same geometry-aware principle while replacing the global attention kernel with a more efficient FLARE-inspired factorisation.

These developments are relevant in the present setting because the full-vehicle design space includes both geometric and thickness variation, while the industrial mesh size makes scalability a practical limitation. Geometry-aware conditioning helps the global processor remain sensitive to spatial configuration, whereas efficient attention helps keep the method tractable at full-vehicle scale.

\subsection{Hybrid local-global architectures}
\label{sec:rw_hybrid}

Hybrid architectures seek to combine the advantages of both paradigms. A local mesh-based stage can preserve neighbourhood-level structure and encode contact-relevant features, whereas a global token-based stage can propagate information efficiently across the full structure. In principle, such a design is attractive for crash simulation, where both local contact-driven events and long-range structural coupling matter.

Recent concurrent work has reinforced this general direction in solid mechanics. In particular, MeshGraphNet-Transformer combines mesh-based local propagation with Transformer-style global processing while preserving a mesh-based graph representation~\citep{iparraguirre2026mgntransformer}. That work explicitly identifies the limited long-range communication of standard MPNNs on large meshes as a central bottleneck and shows the value of combining geometric inductive bias with global attention for impact-type structural dynamics.

However, the relative role of local propagation, global attention, geometry conditioning, and explicit contact modelling remains insufficiently characterised on large industrial crash datasets. This is especially true in a fully autoregressive rollout setting, where local error accumulation, long-range coupling, and dynamic contact activation interact over time. This motivates the systematic comparison carried out in the present work.

\subsection{Crash-surrogate context}
\label{sec:rw_crash}

Recent crash-surrogate studies have shown that both MeshGraphNet-style models and Transolver-style processors are viable building blocks for automotive structural prediction, and that autoregressive rollout is important when the surrogate is intended for deployment rather than one-step forecasting~\citep{nabian2025crashml}. The present work builds on that context, but focuses specifically on full-vehicle lateral pole impact, explicit sparse contact modelling, and downstream safety-relevant evaluation.

Taken together, the literature suggests that local mesh propagation, global token communication, geometry-aware conditioning, and efficient attention are all promising ingredients for learned structural simulation. What remains insufficiently understood is how these ingredients should be combined for large-scale industrial crash rollout, particularly when explicit contact activation and safety-relevant downstream metrics are part of the evaluation. This is the gap addressed in the present study.

\begin{figure}[!t]
    \centering
    \includegraphics[width=\textwidth]{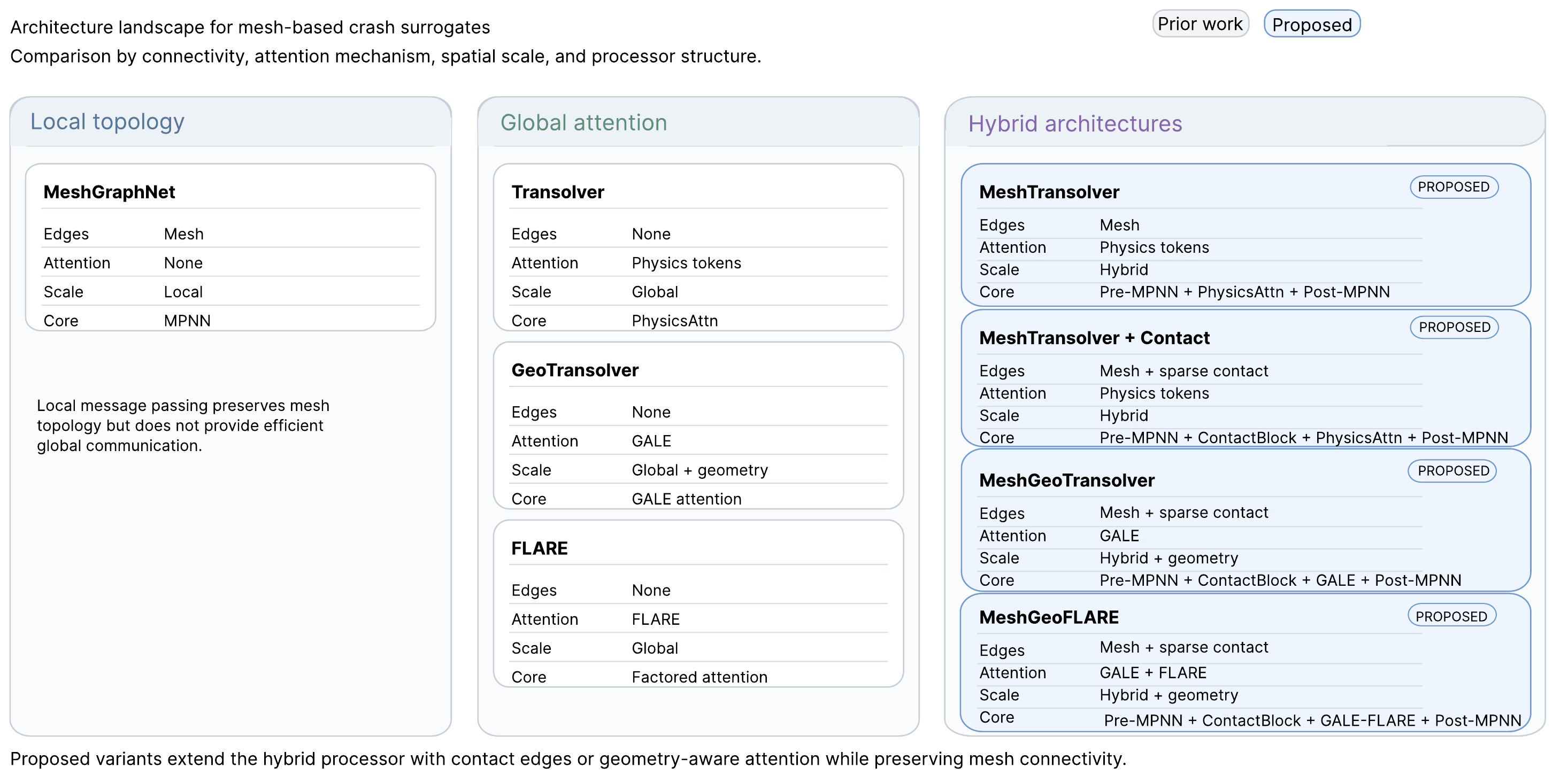}
    \caption{
    Comparison of mesh-based surrogate architectures considered in this work.
    Local MPNN models operate directly on mesh connectivity, global-attention
    models rely on token- or geometry-aware attention without mesh edges, and
    hybrid architectures combine both mechanisms. Proposed variants extend the
    hybrid processor with contact edges or geometry-aware attention while
    preserving mesh connectivity.
    }
    \label{fig:arch_comparison}
\end{figure}

Figure~\ref{fig:arch_comparison} summarises the architectural landscape considered in this work, highlighting differences in connectivity, attention mechanism, spatial scale, and processor design.

%% file: 04_method.tex
\section{Proposed Method}
\label{sec:method}

The proposed architectures are implemented within the NVIDIA PhysicsNeMo framework~\citep{physicsnemo2025}, building on its reference implementations of mesh-based graph neural networks and token-based scientific surrogate models, while extending them with hybrid local--global processors and explicit sparse contact modelling.

\subsection{Overview of the hybrid processor}
\label{sec:overview}

All models follow an encoder--processor--decoder structure. The encoder maps nodal inputs to a latent feature space, the processor updates the latent state through local and/or global operators, and the decoder predicts nodal accelerations used for rollout integration.

For the hybrid variants, the processor is organised as:
\begin{equation}
  \mathbf{H}^{(0)}
  \xrightarrow[\mathcal{E}_{\text{mesh}}]{\text{Pre-MPNN}}
  \mathbf{H}^{(1)}
  \xrightarrow{\text{Global processor}}
  \mathbf{H}^{(2)}
  \xrightarrow[\mathcal{E}_{\text{mesh}}]{\text{Post-MPNN}}
  \mathbf{H}^{(3)},
  \label{eq:hybrid_pipeline}
\end{equation}
where $\mathbf{H}^{(\ell)} \in \mathbb{R}^{N \times d_h}$ denotes the latent nodal representation and $d_h$ is the latent feature dimension.

This decomposition assigns distinct roles to each stage: local feature extraction on the mesh, global structural propagation, and final mesh-based refinement.

\begin{figure}[t]
  \centering
  \includegraphics[width=\linewidth,height=0.85\textheight,keepaspectratio]{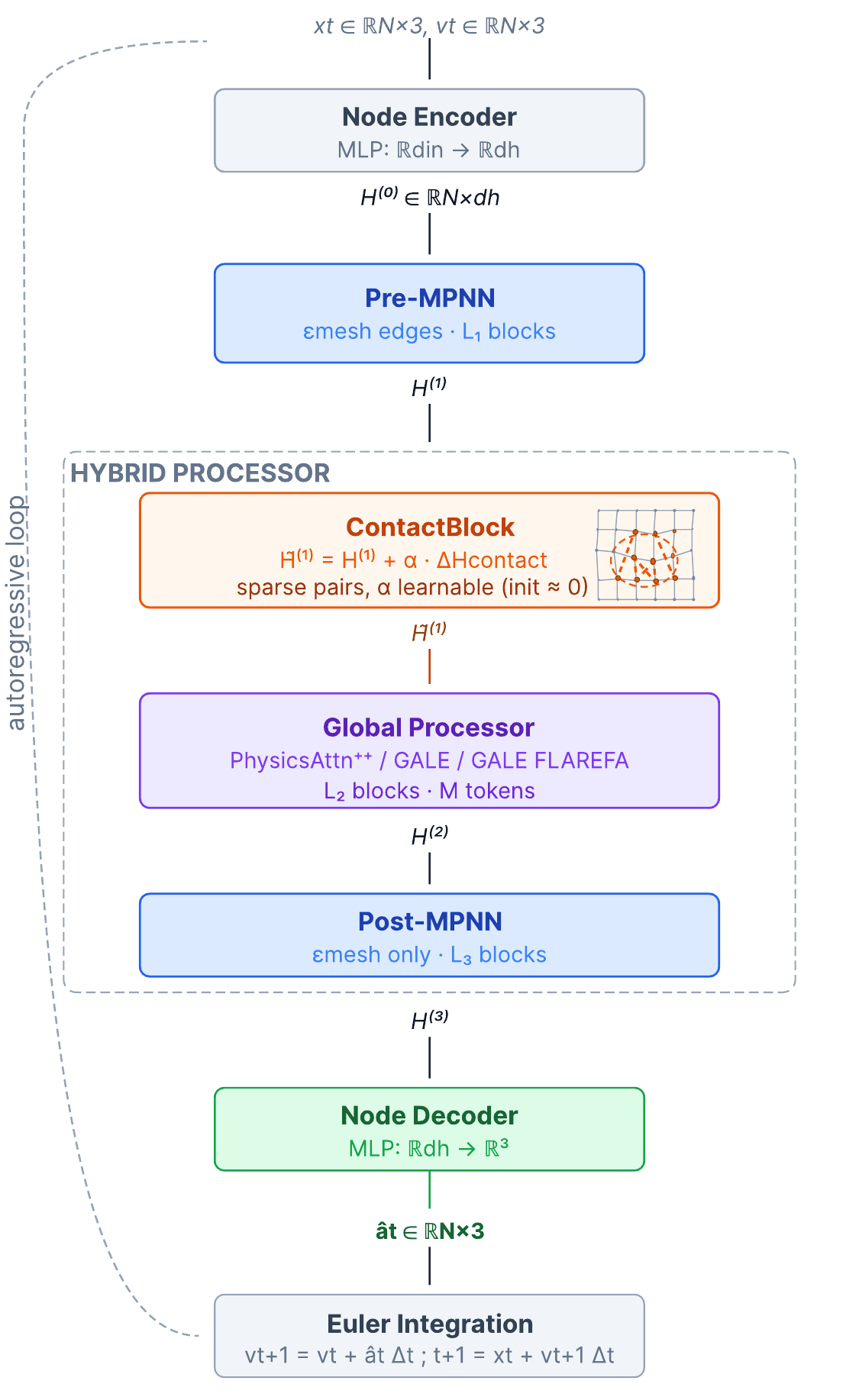}
  \caption{Hybrid processor structure. Local message passing extracts mesh-level features, the global processor propagates long-range interactions, and the final post-MPNN stage remains restricted to the structural mesh.}
  \label{fig:hybrid_processor}
\end{figure}

\subsection{Encoder and local message passing}
\label{sec:encoder}

Each node is embedded into a latent space of dimension $d_h$ using a node encoder MLP. For models that use a mesh graph, local interactions are processed through message-passing neural network (MPNN) blocks operating on $\mathcal{E}_{\text{mesh}}$.

These blocks preserve the FE neighbourhood structure and provide the inductive bias required to capture short-range deformation patterns, boundary effects, and local structural interactions.

\subsection{Global processors}
\label{sec:global_processors}

Three global processors are considered. Transolver-style PhysicsAttention~\citep{wu2024transolver} performs token-based global aggregation through learnable physics slices. The sharpened Transolver$^{++}$ parametrisation~\citep{luo2025transolverpp} is used within the \textit{MeshTransolver} family and is based on the PhysicsNeMo implementation.

GeoTransolver~\citep{adams2025geotransolver} extends this formulation with geometry-aware local embeddings (GALE), making the global processor sensitive to spatial configuration and loading location.

GeoFLARE is introduced in the present work by replacing the GeoTransolver attention kernel with a FLARE-based factorised variant~\citep{puri2025flare}, reducing memory cost while preserving global communication.

The hybrid combination of these global processors with mesh-based propagation and explicit sparse contact modelling is introduced and systematically evaluated in the present work.

\subsection{Generic sparse contact block}
\label{sec:generic_contact}

\begin{figure}[t]
  \centering
  \includegraphics[width=\linewidth]{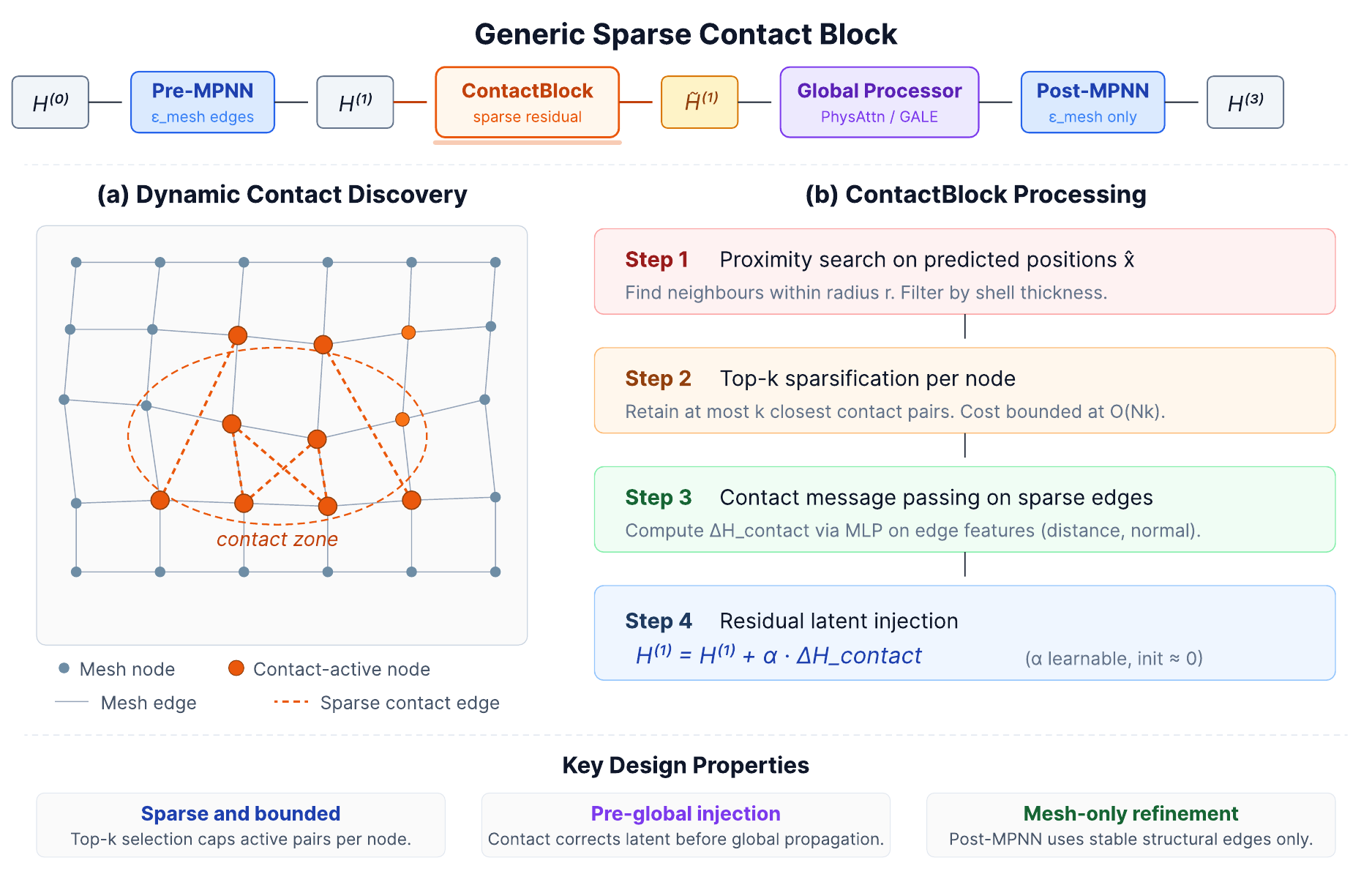}
  \caption{Generic sparse contact block. At each step, a proximity-based radius search identifies candidate interactions from the predicted geometry. These are filtered by thickness and sparsified to the top-$k$ pairs to form the contact graph. The interactions are then converted into a bounded residual latent update and injected after the pre-MPNN and before the global processor. The final refinement stage remains restricted to the mesh.}
  \label{fig:contact_block}
\end{figure}

For \textit{MeshTransolver}, \textit{MeshGeoTransolver}, and \textit{MeshGeoFLARE}, we introduce an explicit \emph{generic sparse contact block}. The processing sequence is summarised in Figure~\ref{fig:contact_block}.

At each rollout step, sparse contact pairs are reconstructed from the current predicted geometry using proximity-based search, thickness-aware filtering, and top-$k$ sparsification. The hybrid processor is modified as:
\begin{equation}
  \mathbf{H}^{(0)}
  \xrightarrow[\mathcal{E}_{\text{mesh}}]{\text{Pre-MPNN}}
  \mathbf{H}^{(1)}
  \xrightarrow{\text{ContactBlock}}
  \widetilde{\mathbf{H}}^{(1)}
  \xrightarrow{\text{Global processor}}
  \mathbf{H}^{(2)}
  \xrightarrow[\mathcal{E}_{\text{mesh}}]{\text{Post-MPNN}}
  \mathbf{H}^{(3)},
  \label{eq:contact_pipeline}
\end{equation}

The ContactBlock computes a residual latent correction:
\begin{equation}
  \widetilde{\mathbf{H}}^{(1)} =
  \mathbf{H}^{(1)} + \alpha \,\Delta \mathbf{H}_{\text{contact}},
\end{equation}
where $\Delta \mathbf{H}_{\text{contact}}$ is computed from sparse contact pairs and $\alpha$ is a learnable scalar initialised near zero.

This formulation introduces contact explicitly while keeping the representation sparse and bounded. The correction is injected before global propagation, and the final refinement stage remains restricted to the structural mesh.

Overall, the design captures informative contact interactions while preventing rapidly changing proximity relations from destabilising the final mesh-level refinement stage.

\subsection{Model variants}
\label{sec:variants}

Table~\ref{tab:model_variants} summarises the main architectural and hyperparameter choices. As introduced in previous sections, $d_h$ represents the latent feature dimension, $M$ defines the number of global tokens used for long-range communication, and $k$ controls the maximum number of active contact pairs per node.
The column $L$ blocks indicates the number of processing layers in each stage.
For hybrid models, it is reported as Pre-MPNN + Global + Post-MPNN.

The column $M$ tokens denotes the number of latent tokens used by the global processor
(PhysicsAttention, GALE, or GALE\_FA), which controls the capacity of global information aggregation.

The Contact column indicates whether the generic sparse contact block is used.
When enabled, the contact configuration specifies the maximum number of active contact pairs per node ($k$),
which bounds the computational cost of contact processing.

Geometry-aware models (GeoTransolver and GeoFLARE variants) use geometry conditioning (GALE),
which augments token assignment with spatial embeddings derived from nodal positions.

\textit{MeshTransolver} combines pre-MPNN, PhysicsAttention$^{++}$, and post-MPNN stages. Contact-aware variants incorporate the generic sparse contact block between local and global processing.

\textit{MeshGeoTransolver} and \textit{MeshGeoFLARE} preserve the same hybrid structure while replacing the global processor with GALE and GALE\_FA, respectively.
This parameterisation highlights the trade-off between local depth, global capacity,
and contact sparsity across the evaluated architectures.

\begin{table}[t]
\centering
\caption{Hyperparameter summary for all evaluated architectures. $d_{\text{model}} = 128$ for all models. Contact config indicates the sparse-contact settings used in the autoregressive hybrid variants.}
\label{tab:model_variants}
\footnotesize
\resizebox{\linewidth}{!}{%
\begin{tabular}{@{}llcccc@{}}
\toprule
Model & Processor & $L$ blocks & $M$ tokens & Contact & Contact config \\
\midrule
MeshGraphNet & Local MPNN & 6 & --- & No & --- \\
Transolver & PhysicsAttn & 6 & 128 & No & --- \\
\textit{MeshTransolver} & Pre-MPNN + PhysicsAttn$^{++}$ + Post-MPNN & 1+6+2 & 128 & No & --- \\
\textit{MeshTransolver+Contact} & Pre-MPNN + ContactBlock + PhysicsAttn$^{++}$ + Post-MPNN & 1+6+2 & 128 & Yes & generic sparse block, $k{=}32$ \\
GeoTransolver & GALE attention & 4 & 128 & No & --- \\
GeoFLARE & GALE\_FA attention & 4 & 128 & No & --- \\
\textit{MeshGeoFLARE} & Pre-MPNN + ContactBlock + GALE\_FA + Post-MPNN & 1+4+2 & 128 & Yes & generic sparse block, $k{=}16$ \\
\textit{MeshGeoTransolver} & Pre-MPNN + ContactBlock + GALE + Post-MPNN & 1+4+2 & 128 & Yes & generic sparse block, $k{=}16$ \\
\bottomrule
\end{tabular}%
}
\end{table}

\subsection{Autoregressive rollout}
\label{sec:rollout}

All models are trained and evaluated in closed-loop autoregressive mode. At each timestep, the model predicts nodal accelerations:
\begin{equation}
  \mathbf{a}_t = f_\theta(\mathbf{x}_t, \mathbf{v}_t),
\end{equation}
which are integrated as:
\begin{align}
  \mathbf{v}_{t+1} &= \mathbf{v}_t + \mathbf{a}_t \Delta t, \\
  \mathbf{x}_{t+1} &= \mathbf{x}_t + \mathbf{v}_{t+1} \Delta t.
\end{align}

The predicted state is fed back into the next step. In the contact-aware variants, sparse contact is recomputed from the predicted geometry at each rollout step.
\begin{figure}[t]
  \centering
  \includegraphics[width=\linewidth]{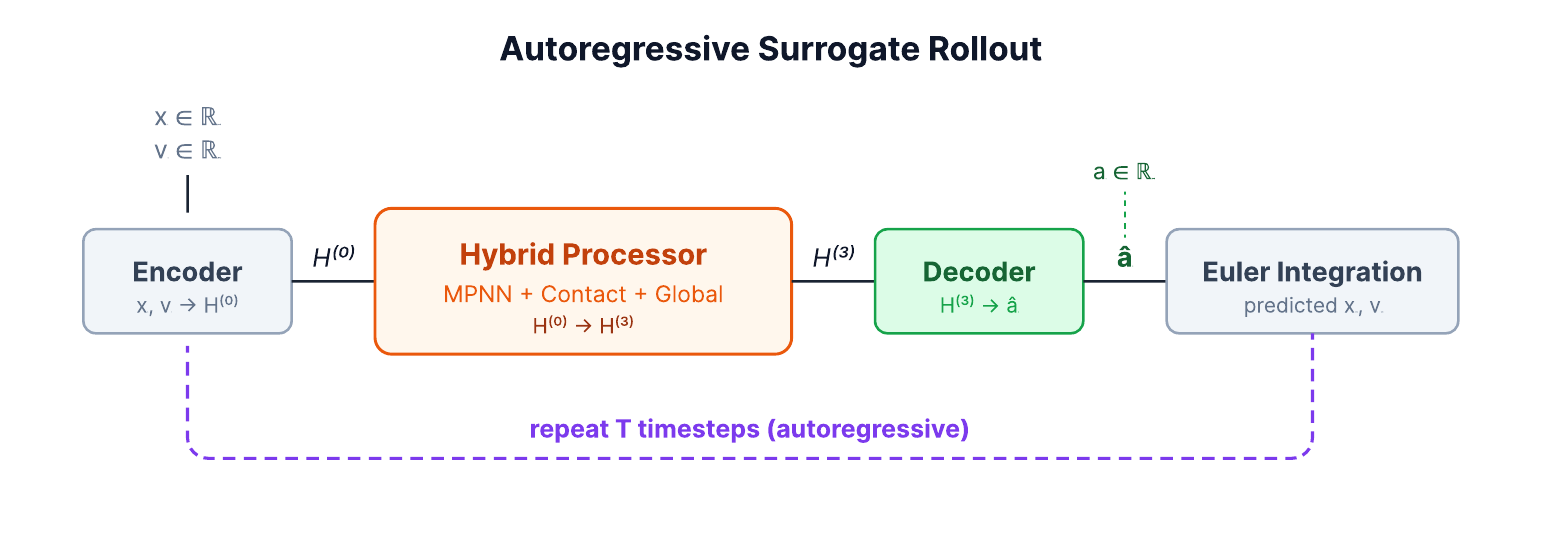}
  \caption{Autoregressive rollout. The model predicts acceleration from the current state, integrates forward in time, and feeds the predicted configuration back into the next step. Contact-aware variants recompute sparse contact at each step.}
  \label{fig:rollout}
\end{figure}

%% file: 05_experiments.tex
\section{Experimental Setup}
\label{sec:experiments}

\textit{Disclaimer: Due to confidentiality restrictions, absolute numerical values for proprietary geometric parameters and specific performance metrics have been normalized in the figures.}

\subsection{Full-vehicle dataset}
\label{sec:dataset}

The case study consists of explicit finite-element crash simulations of a parameterised full-vehicle model under a lateral pole-impact load case aligned with the Euro NCAP \citep{euroncap2023} pole-impact setup. The vehicle strikes a rigid cylindrical pole of 254~mm diameter at 33~km/h. This slightly exceeds the 32~km/h nominal speed specified by UN R135 and reflects a conservative internal validation protocol chosen to cover tolerance margins around the regulatory condition. The pole-impact reference configuration follows the standard oblique side-impact arrangement, with the impact reference line defined at 75$^\circ$ to the vehicle longitudinal centreline.

\begin{figure}[t]
  \centering
  \includegraphics[width=\linewidth]{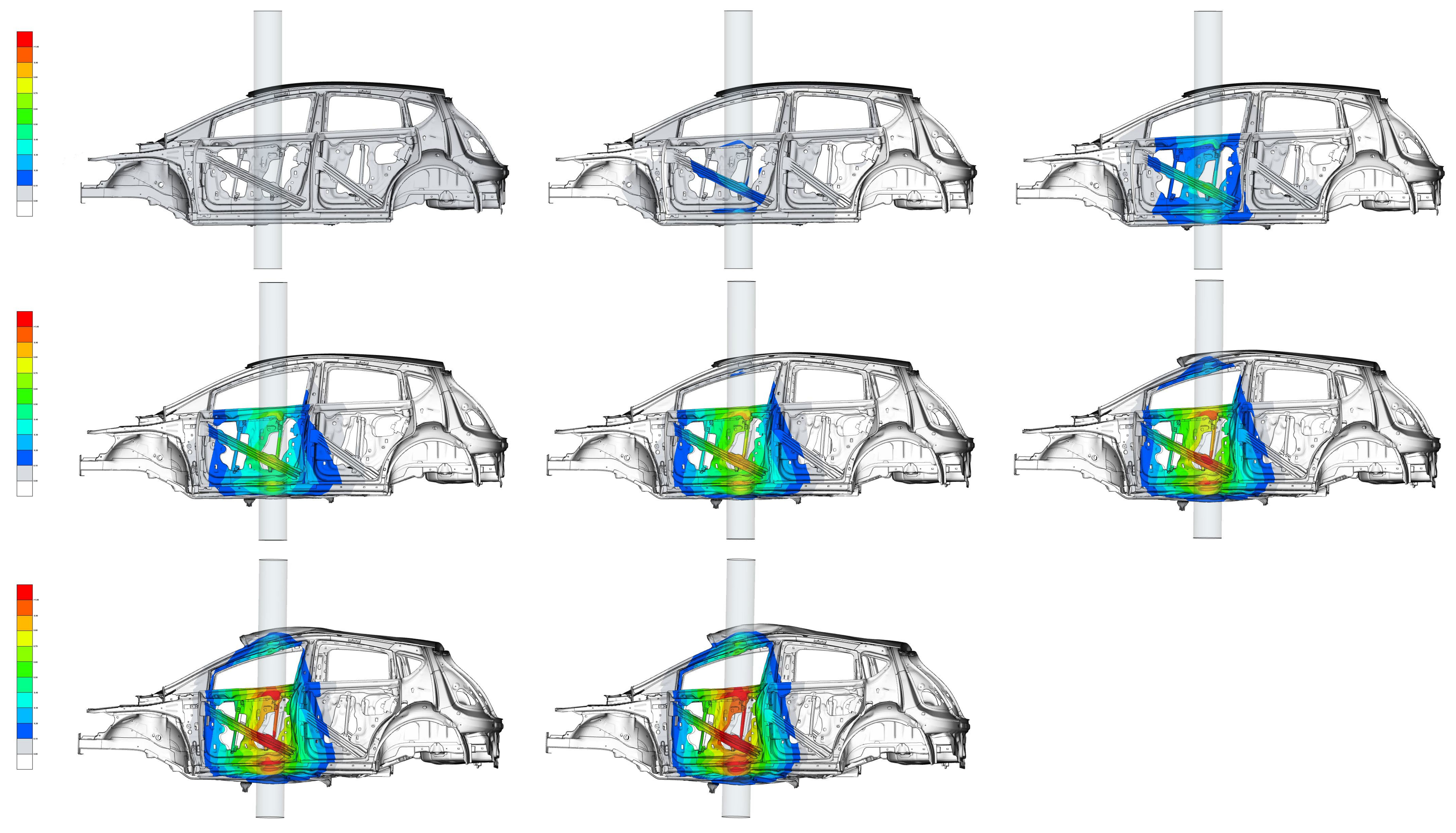}
  \caption{Temporal evolution of the lateral pole-impact benchmark at representative timesteps. The results correspond to the nominal (standard) values of the DOE design space. The sequence illustrates the progressive intrusion of the side structure and the global redistribution of deformation across the structure. The displacement fringe is normalized.}
  \label{fig:crash_sequence}
\end{figure}

The structural response is recorded over 70~ms with outputs every 5~ms ($T=15$). Figure~\ref{fig:crash_sequence} illustrates the temporal evolution of the benchmark.

The underlying FE model is a full-vehicle crash model including the main body structure and vehicle subsystems typically present in an industrial side-impact setup. However, the surrogate prediction target is restricted to the structural components retained in the learning graph. In other words, the complete simulation is used to generate the crash response, but only the structural domain is predicted by the neural surrogate after preprocessing and graph construction. Material rupture or tearing is not modelled in the present benchmark; the study focuses on large-deformation structural dynamics with contact.

After preprocessing and graph construction, the learning problem involves approximately 280{,}000 structural nodes per simulation.

\subsection{Parameterised design space}
\label{sec:dataset_generation}

The dataset is generated with Latin Hypercube Sampling (LHS)~\citep{mckay1979comparison} over an eight-dimensional design space combining crash kinematics, sheet-thickness variation, and geometric morphing. LHS is used because it provides broad coverage of the parameter space under a limited simulation budget, which is preferable to simple random sampling when building a representative DOE for expensive vehicle crash simulations. This LHS-derived DOE defines the sampled parameter space used throughout training, validation, and test split construction.

\begin{figure}[!t]
  \centering
  \includegraphics[width=\linewidth]{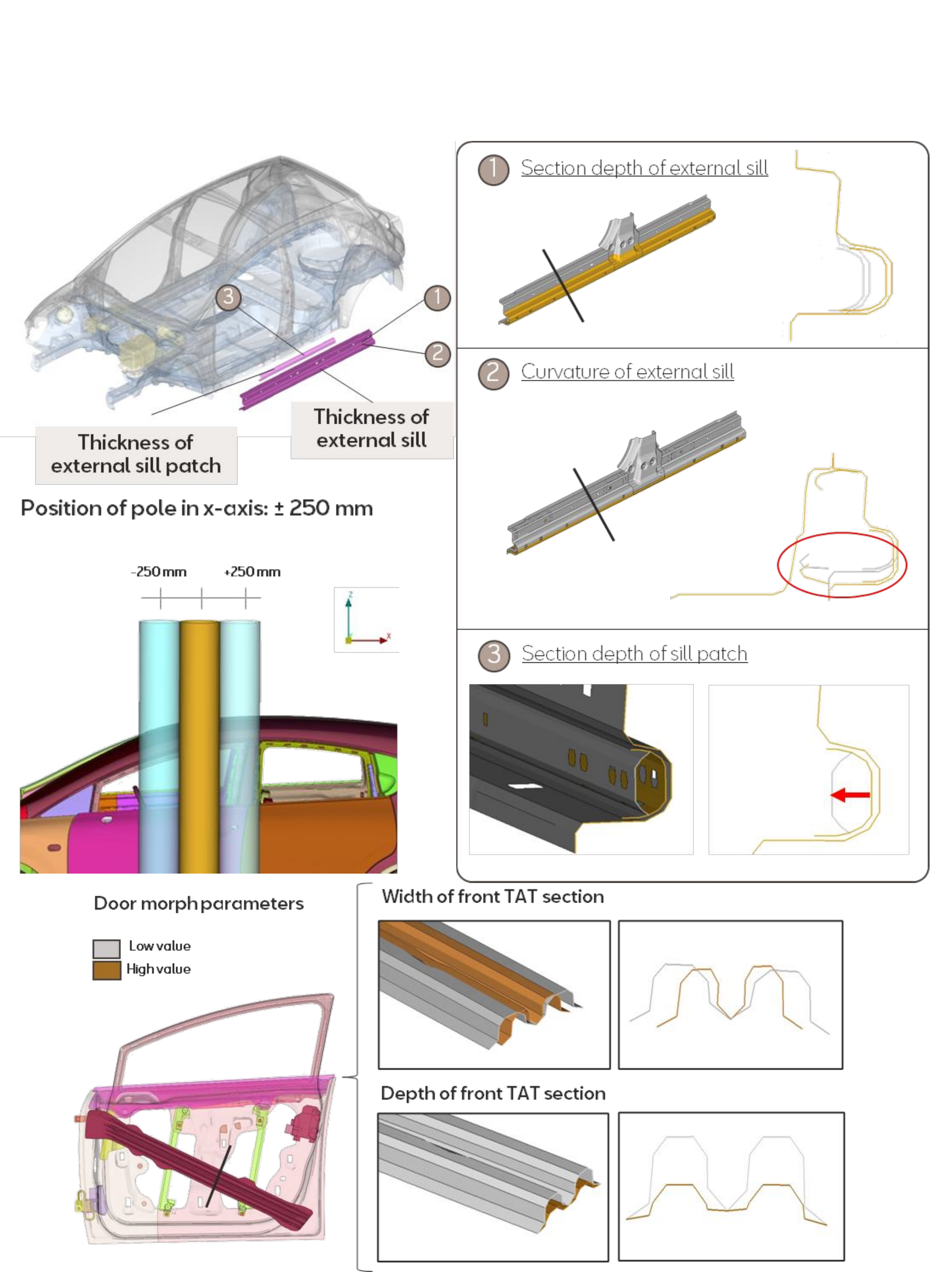}
  \caption{Parameterised full-vehicle model used in the lateral pole-impact benchmark. The DOE includes thickness, geometry, and pole-position variation. Geometric labels are normalized.}
  \label{fig:se350_params}
\end{figure}

Table~\ref{tab:se350_params} lists the parameters and their ranges, while Figure~\ref{fig:se350_params} shows the corresponding geometric regions.

\begin{table}[t]
\centering
\caption{Parameterised design space (8 LHS variables). \textit{Standard}: nominal production value. \textit{Low/High}: sampling bounds.}
\label{tab:se350_params}
\begin{tabular}{@{}lrrrl@{}}
\toprule
Parameter & Standard & Low & High & Unit \\
\midrule
Pole position ($x$-axis)        & 1203 &  953 & 1453 & mm \\
Thickness, external sill        &  1.5 &  1.2 &  1.8 & mm \\
Thickness, external sill patch  &  1.2 &  0.8 &  1.6 & mm \\
Depth section, front TAT        &    0 &   $-2$ &   12 & mm \\
Width, front TAT                &    0 &   $-7$ &    7 & mm \\
Section depth, external sill    &    0 &  $-10$ &   10 & mm \\
Radius curvature, external sill &    0 &  $-10$ &    5 & mm \\
Section depth, sill patch       &    0 &  $-25$ &    0 & mm \\
\bottomrule
\end{tabular}
\end{table}

\subsection{DOE-aware split validation}
\label{sec:dataset_splits}

The dataset contains 200 simulations, split into 150 training, 25 validation, and 25 test samples. The split is not random: it is constructed to preserve coverage of the DOE while avoiding pathological concentration of outliers in the test set. This is important because a biased split can make the benchmark either unrealistically easy or artificially pessimistic.

\begin{figure}[!t]
  \centering
  \includegraphics[width=\linewidth]{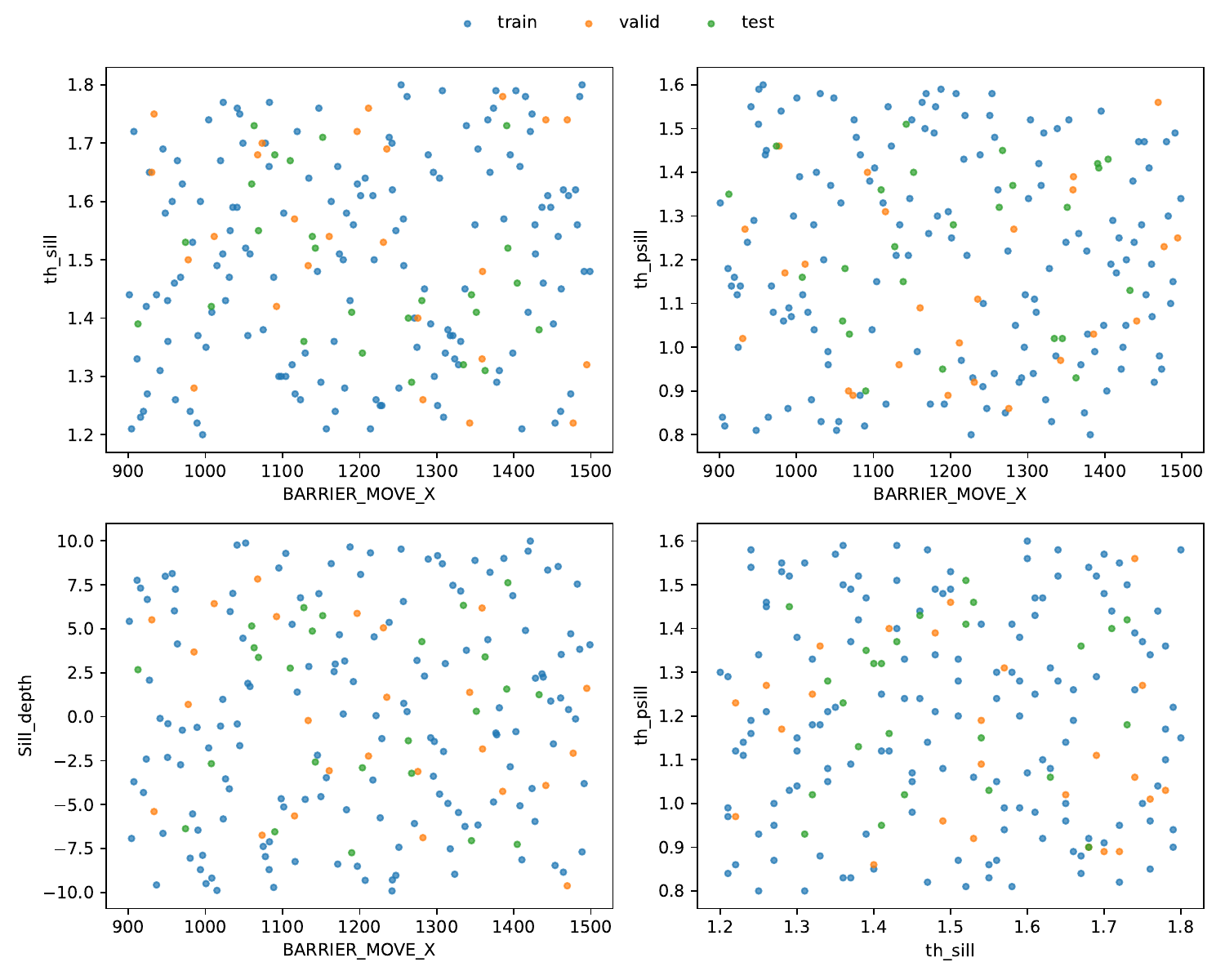}
  \caption{Pairwise distribution of selected design variables in the LHS design space. Training (blue), validation (orange), and test (green) splits exhibit consistent coverage across loading, thickness, and geometric parameters. Diagonal panels show marginal distributions, whereas off-diagonal panels illustrate joint coverage. For clarity, only a representative subset of variables is shown. Distribution axes are normalized.}
  \label{fig:split_scatter}
\end{figure}

To assess representativeness, we validate the split using complementary distributional diagnostics over the design variables, including Wasserstein distance, Kolmogorov--Smirnov statistics, and coverage-oriented criteria in low-dimensional projections. Figure~\ref{fig:split_scatter} illustrates the resulting overlap between training, validation, and test subsets in representative pairwise projections of the design space.

\subsection{Training details}
\label{sec:training_setup}

All architectures were evaluated under a common training and hyperparameter configuration to ensure a controlled comparison. This isolates the effect of the architectural components and establishes a fair shared baseline, although it formally leaves open the possibility that some configurations could be further optimized with model-specific tuning. The reported rankings should therefore be interpreted under this shared configuration rather than as the result of exhaustive per-model hyperparameter search.

We use AdamW~\citep{loshchilov2019decoupled} with cosine annealing~\citep{loshchilov2017sgdr}, an initial learning rate of $10^{-4}$, mixed bfloat16 precision, activation checkpointing, and batch size 1 (one full rollout per optimisation step). Early stopping is applied on validation mean-squared error with a patience of approximately 15--20 epochs, and training is performed on 8 $\times$ NVIDIA H100 GPUs.

\subsection{Compared models}
\label{sec:model_families}

We evaluate eight architectures spanning three families: local mesh-based models, global attention models, and hybrid local--global variants. Their detailed definitions were introduced in Section~\ref{sec:method}, the main architecture hyperparameters are summarised in Table~\ref{tab:hyperparams}, and the corresponding model sizes and GPU memory consumption are summarised in Table~\ref{tab:model_capacity}.

Table~\ref{tab:hyperparams} reports the main architectural settings. For hybrid models, $L_{\mathrm{pre}}$, $L_{\mathrm{attn}}$, and $L_{\mathrm{post}}$ denote the number of local pre-processing, global-processing, and mesh-only refinement blocks, respectively. $M$ denotes the number of latent tokens or slices used by the global processor, $d_h$ is the hidden dimension, and $D_{\mathrm{node}}$ is the input node-feature dimension before encoding. The Contact column indicates whether the generic sparse contact block is enabled.

\begin{table}[t]
\centering
\caption{Architecture hyperparameters per model. $L_{\mathrm{pre}}$/$L_{\mathrm{post}}$: pre/post MPNN blocks. $L_{\mathrm{attn}}$: attention blocks. $M$: number of tokens or slices. $d_h$: hidden dimension. $D_{\mathrm{node}}$: input node-feature dimension before encoding.}
\label{tab:hyperparams}
\footnotesize
\resizebox{\linewidth}{!}{%
\begin{tabular}{@{}llccccccc@{}}
\toprule
Model & Attention & $L_{\mathrm{pre}}$ & $L_{\mathrm{attn}}$ & $L_{\mathrm{post}}$ & $M$ & $d_h$ & $D_{\mathrm{node}}$ & Contact \\
\midrule
MGN & --- & \multicolumn{3}{c}{$L=6$ MPNN} & --- & 128 & 7 & No \\
Transolver & PhysicsAttention & --- & 6 & --- & 128 & 128 & 5 & No \\
\textit{MeshTransolver} & PhysicsAttention$^{++}$ & 1 & 6 & 2 & 128 & 128 & 11 & No \\
\textit{MeshTransolver+Contact} & PhysicsAttention$^{++}$ & 1 & 6 & 2 & 128 & 128 & 11 & Yes ($k=32$) \\
GeoTransolver & GALE & --- & 4 & --- & 128 & 128 & 5 & No \\
GeoFLARE & GALE\_FA & --- & 4 & --- & 128 & 128 & 5 & No \\
\textit{MeshGeoFLARE} & GALE\_FA & 1 & 4 & 2 & 128 & 128 & 11 & Yes ($k=16$) \\
\textit{MeshGeoTransolver} & GALE & 1 & 4 & 2 & 128 & 128 & 11 & Yes ($k=16$) \\
\bottomrule
\end{tabular}%
}
\end{table}

\begin{table}[!t]
\centering
\caption{Representative model capacity and training duration for the configurations used in the reported full-vehicle comparison. Parameter counts are taken from the trained configurations summarised in the experiment logs. Epoch time corresponds to the average wall-clock time per training epoch for the reported configuration.}
\label{tab:model_capacity}
\footnotesize
\begin{tabular}{@{}lccc@{}}
\toprule
Model & Parameters & Max.memory per GPU [GB] & Time/epoch [s] \\
\midrule
MeshGraphNet & 994.5K & 17.56 & 53.44 \\
Transolver & 774.6K & 24.55 & 95.11 \\
\textit{MeshTransolver} & 1.30M & 49.90 & 205.0 \\
\textit{MeshTransolver+Contact} & 1.32M & 72.80 & 246.86 \\
GeoTransolver & 1.72M & 35.82 & 222.63 \\
GeoFLARE & 1.66M & 42.57 & 206.74 \\
\textit{MeshGeoFLARE} & 1.45M & 76.04 & 155.89 \\
\textit{MeshGeoTransolver} & 1.41M & 44.38 & 297.63 \\
\bottomrule
\end{tabular}
\end{table}

\subsection{Evaluation protocol}
\label{sec:eval_protocol}

All reported metrics correspond to the 25-sample in-distribution test set under full autoregressive rollout over the complete 70~ms horizon. We report temporal mean RMSE, final-step RMSE, and final-step relative RMSE. We additionally report occupant survival-space error because deployment relevance is not fully captured by global nodal error alone. For contact-aware models, sparse contact is rebuilt from \emph{predicted} positions at rollout time, ensuring that evaluation matches the intended deployment setting.

%% file: 06_results.tex
\section{Results: Full-Vehicle Crash Benchmark}
\label{sec:results}

Results are reported on the 25-sample in-distribution test set evaluated under fully autoregressive rollout. The displacement root-mean-square error (RMSE) for a given timestep $t$ and test sample $k$ is defined as:
\begin{equation}
\mathrm{RMSE}_t^{(k)} = \sqrt{\frac{1}{N} \sum_{i=1}^N \left\| \mathbf{u}_{i,t}^{\mathrm{pred}(k)} - \mathbf{u}_{i,t}^{\mathrm{GT}(k)} \right\|_2^2},
\end{equation}
where $N$ is the number of nodes, and $\mathbf{u}_{i,t}$ is the nodal displacement vector.

From this, we define the primary metric, the \emph{temporal mean RMSE} ($\mathrm{RMSE}_{\mu}$), as the average over all $T$ rollout steps and all $K$ test samples:
\begin{equation}
\mathrm{RMSE}_{\mu} = \frac{1}{K} \sum_{k=1}^K \left( \frac{1}{T} \sum_{t=1}^T \mathrm{RMSE}_t^{(k)} \right).
\end{equation}

Similarly, the \emph{final-step relative RMSE} (Rel.~RMSE) normalises the final-step error by the ground-truth displacement magnitude at $t=T$, averaged over the test set:
\begin{equation}
\text{Rel.~RMSE} = \frac{1}{K} \sum_{k=1}^K \frac{\mathrm{RMSE}_T^{(k)}}{\sqrt{\frac{1}{N}\sum_{i=1}^N \left\| \mathbf{u}_{i,T}^{\mathrm{GT}(k)} \right\|_2^2}}.
\end{equation}

\subsection{In-distribution accuracy}
\label{sec:results_id}

Table~\ref{tab:id_results} summarises the global field accuracy for all eight evaluated architectures. Note that metric aggregation formats are tailored to their purpose: $\mathrm{RMSE}_{\mu}$ and Rel.~RMSE are reported as single dataset-wide mean values to provide concise overall rankings, whereas the absolute final-step error ($\mathrm{RMSE}_{\mathrm{final}}$) is reported as the test-set mean $\pm$ standard deviation to explicitly illustrate the dispersion of absolute predictions at the most challenging point of the rollout.

\textit{MeshGeoFLARE} attains the lowest error among all models ($\mathrm{RMSE}_{\mu}=3.20$~mm, final-step relative error $=0.0137$), followed by \textit{MeshGeoTransolver} and \textit{GeoFLARE}. This ranking suggests that while geometry-aware global attention is effective, combining it with local mesh-based propagation in a hybrid design provides the strongest quantitative accuracy for the present benchmark.

Pure geometry-conditioned attention models (GeoTransolver and GeoFLARE) achieve competitive temporal mean errors ($\mathrm{RMSE}_{\mu}$ between 3.8 and 4.2~mm), indicating that geometry-aware global attention can capture the dominant global crash kinematics under the shared hyperparameter configuration. However, as discussed in Section~\ref{sec:results_qualitative}, these models can exhibit higher local spatial noise compared to the hybrid variants. The comparison therefore suggests that, for this benchmark and configuration, the best quantitative performance is obtained by combining mesh-based local propagation with a global processor, with additional gains from geometry-aware global aggregation.

\begin{table}[ht]
\centering
\caption{In-distribution test results on the 25-sample full-vehicle test set. $\mathrm{RMSE}_{\mu}$ denotes the temporal mean over the 15 rollout steps, and $\mathrm{RMSE}_{\mathrm{final}}$ denotes the final timestep at 70~ms. Bold indicates the best value in each column.}
\label{tab:id_results}
\footnotesize
\resizebox{\linewidth}{!}{%
\begin{tabular}{@{}lccc@{}}
\toprule
Model & $\mathrm{RMSE}_{\mu}$ (mm) & $\mathrm{RMSE}_{\mathrm{final}}$ (mm) & Rel.~RMSE \\
\midrule
\textit{MeshGeoFLARE} & $\mathbf{3.20}$ & $\mathbf{5.95 \pm 1.34}$ & \textbf{0.0137} \\
\textit{MeshGeoTransolver} & $3.43$ & $5.98 \pm 1.24$ & \textbf{0.0137} \\
GeoFLARE & $3.88$ & $6.89 \pm 1.14$ & $0.0158$ \\
GeoTransolver & $4.12$ & $7.40 \pm 1.13$ & $0.0170$ \\
\textit{MeshTransolver+Contact} & $4.13$ & $7.92 \pm 2.20$ & $0.0183$ \\
Transolver & $4.88$ & $9.15 \pm 2.07$ & $0.0211$ \\
\textit{MeshTransolver} & $13.60$ & $27.92 \pm 10.41$ & $0.0637$ \\
MeshGraphNet & $16.23$ & $32.90 \pm 12.39$ & $0.0756$ \\
\bottomrule
\end{tabular}%
}
\end{table}

\begin{figure}[!t]
\centering
\includegraphics[width=\linewidth]{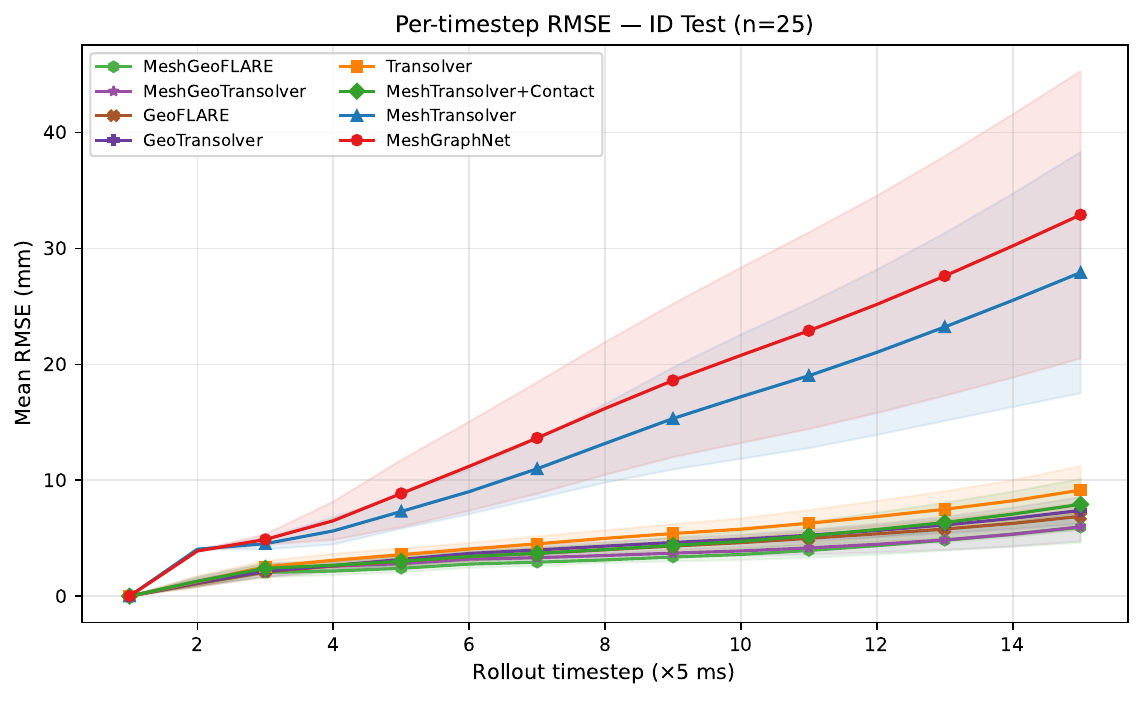}
\caption{Per-timestep RMSE for selected local, global, and hybrid models. Hybrid models exhibit the most stable error growth over the 70~ms rollout horizon.}
\label{fig:rmse_hybrid}
\end{figure}

Figure~\ref{fig:rmse_hybrid} shows that hybrid architectures accumulate error more gradually and remain much more stable over time.

\subsection{Temporal behaviour and effect of contact modelling}
\label{sec:results_contact}

Within the Hybrid mesh–attention models, the most consequential architectural change is the introduction of explicit sparse contact modelling. Comparing the contact-free \textit{MeshTransolver} with the contact-aware variant reduces $\mathrm{RMSE}_{\mu}$ from 13.60 to 4.13~mm, corresponding to a $3.29\times$ improvement.

\begin{figure}[!t]
  \centering
  \includegraphics[width=\linewidth]{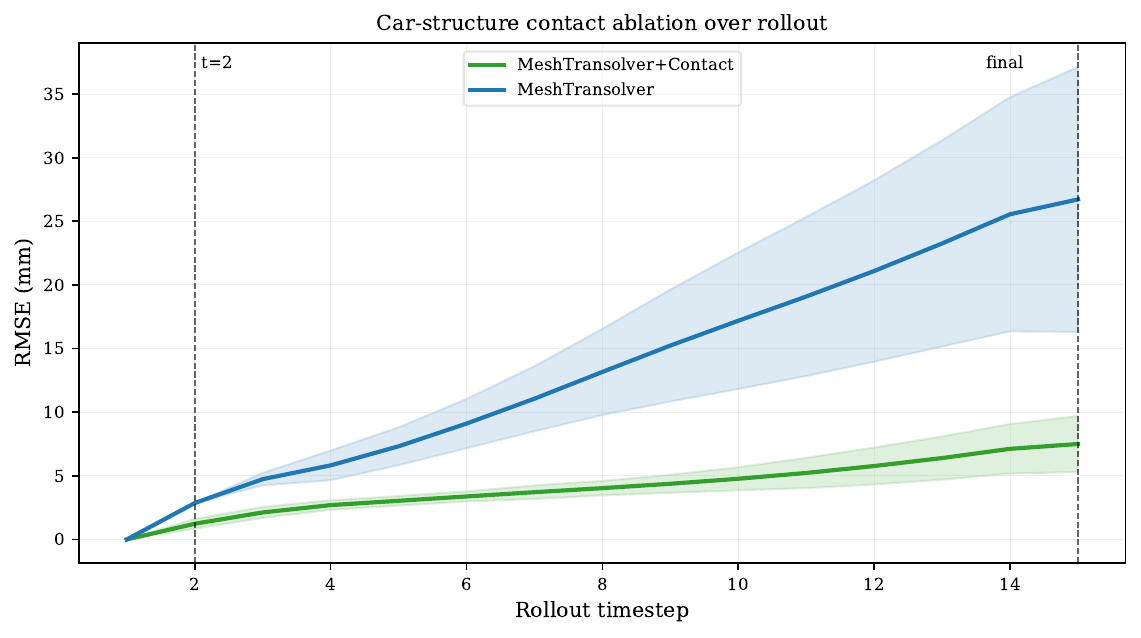}
  \caption{Temporal effect of explicit sparse contact modelling within the \textit{MeshTransolver} family. The largest gain appears during the early-impact regime, where dynamic interactions first become structurally relevant.}
  \label{fig:contact_ablation_temporal}
\end{figure}

Figure~\ref{fig:contact_ablation_temporal} makes this temporal behaviour explicit. The largest benefit appears when contact first becomes dynamically important, which is consistent with the intended role of the generic sparse contact block: to inject interaction information that is absent from the static mesh before the global processor propagates the disturbance through the structure.

A useful interpretation is to assign separate roles to the three stages of the contact-aware hybrid processor. The pre-MPNN stage extracts a locally structured latent state on the mesh, the sparse contact block injects contact-dependent corrections into that representation, and the global processor propagates their effect across the structure. The final post-MPNN stage then refines the deformation on the stable structural mesh only. This interpretation is consistent with the implementation described in Section~\ref{sec:generic_contact} and does not require dynamic contact pairs to remain active during the final refinement stage.

\subsection{Qualitative deformation examples}
\label{sec:results_qualitative}

\begin{figure*}[!t]
  \centering
  \includegraphics[width=\textwidth,height=0.85\textheight,keepaspectratio]{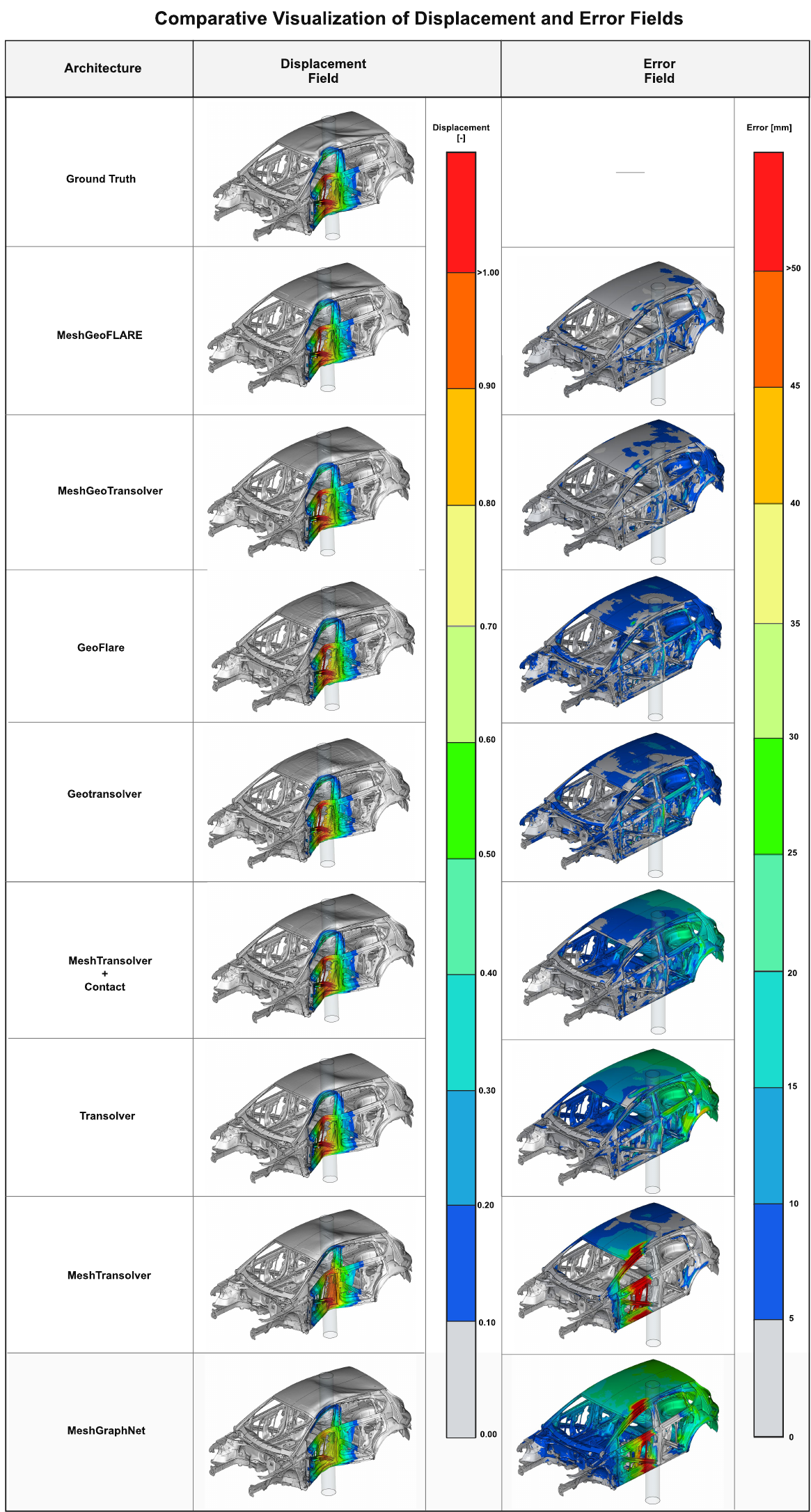}
  \caption{Ground-truth and predicted deformation fields at the final timestep (70~ms) for a representative test case. GeoTransolver and GeoFLARE are omitted for compactness, and are instead compared in the side-view analysis of Figure~\ref{fig:side_view_field_regularization}. The shown hybrid models reproduce both global and local deformation patterns faithfully. The displacement field and error fringes are normalized to the maximum observed range.}
  \label{fig:deformation_comparison}
\end{figure*}

The quantitative ranking is consistent with the qualitative behaviour of the predicted deformation fields. Figure~\ref{fig:deformation_comparison} provides a representative comparison at the final timestep ($t = 70$ ms), showing both the predicted displacement fields and the corresponding error fields with respect to the ground truth.

The error field is defined as the nodal Euclidean displacement difference:
\[
e_i = \left\| \mathbf{u}_i^{\mathrm{pred}} - \mathbf{u}_i^{\mathrm{GT}} \right\|_2,
\]
where $\mathbf{u}_i$ denotes the displacement vector at node $i$. This definition captures both magnitude and directional discrepancies in the predicted deformation.

\begin{figure}[ht]
  \centering
  \begin{subfigure}[b]{0.48\linewidth}
    \includegraphics[width=\linewidth]{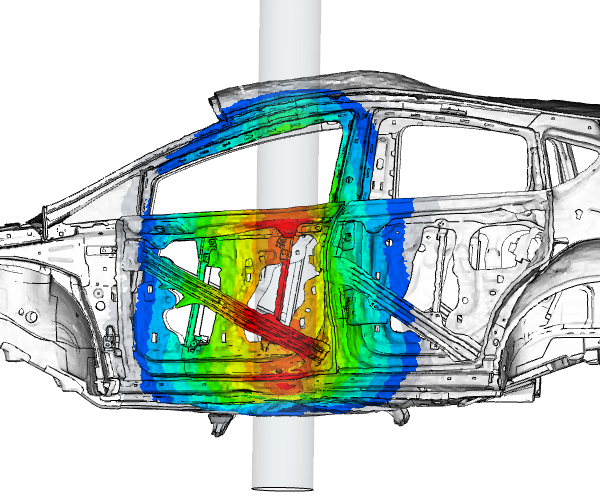}
    \caption{GeoTransolver}
  \end{subfigure}
  \hfill
  \begin{subfigure}[b]{0.48\linewidth}
    \includegraphics[width=\linewidth]{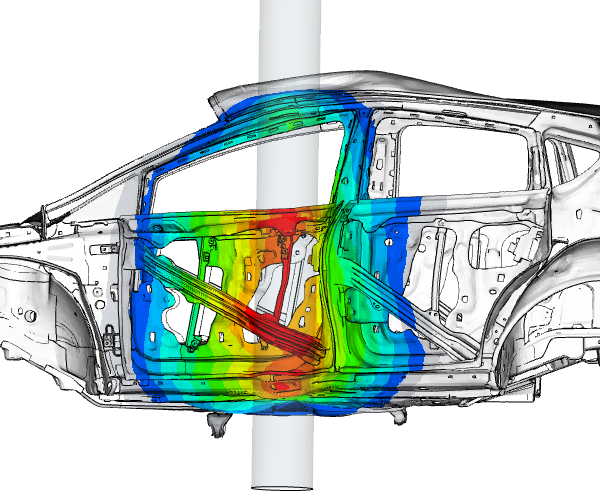}
    \caption{MeshGeoTransolver}
  \end{subfigure}
  \caption{Zoomed side-view comparison of the predicted deformation field at 70~ms. The pure attention model (GeoTransolver) achieves competitive scalar error but shows higher local spatial noise (irregularities). The hybrid variant (MeshGeoTransolver) better preserves field regularity and structural plausibility. Displacement fields are normalized.}
  \label{fig:side_view_field_regularization}
\end{figure}

To ensure a fair visual comparison across architectures, all displacement fields share a common color scale, and all error fields share a separate common scale. This avoids misleading visual differences caused by automatic rescaling and allows consistent assessment of spatial error patterns. GeoTransolver and GeoFLARE are omitted from this qualitative panel to focus the comparison on models that preserve high field regularity; although they reach competitive quantitative RMSE, their displacement fields exhibit localized spatial irregularities that are further analyzed in Figure~\ref{fig:side_view_field_regularization}.

The strongest hybrid models preserve both the global structural motion and the local deformation patterns, including sharp folding regions and high-gradient zones near contact interfaces.
At the same time, geometry-only attention models can achieve competitive global error metrics, but their displacement fields may exhibit localised spatial irregularities and higher noise in some structural regions, as analysed further in Figure~\ref{fig:side_view_field_regularization}. In the error fields, hybrid models exhibit more localised and lower-magnitude error distributions, consistent with their stronger local inductive bias.

A critical industrial observation is that scalar RMSE alone does not fully characterize the usability of a crash surrogate. Although GeoTransolver and GeoFLARE are quantitatively competitive, side-view inspection reveals higher levels of spatial noise and local deformation irregularities in the predicted displacement field. In contrast, MeshGeoTransolver and MeshGeoFLARE combine competitive scalar accuracy with smoother and more physically interpretable deformation fields. This highlights the value of combining mesh-local propagation with geometry-aware global attention to ensure both quantitative precision and structural plausibility.

\subsection{Occupant survival space}
\label{sec:results_safety}

Occupant survival space is used here as a downstream evaluation metric rather than as a training target. It measures whether the surrogate preserves a safety-relevant quantity derived from the predicted structural response. In this context, the sign of the error matters: positive values indicate that the surrogate predicts a larger residual survival distance than the FE reference (non-conservative underestimation of intrusion), whereas negative values correspond to a conservative overestimation.

\begin{figure}[t]
\centering
\includegraphics[width=\linewidth]{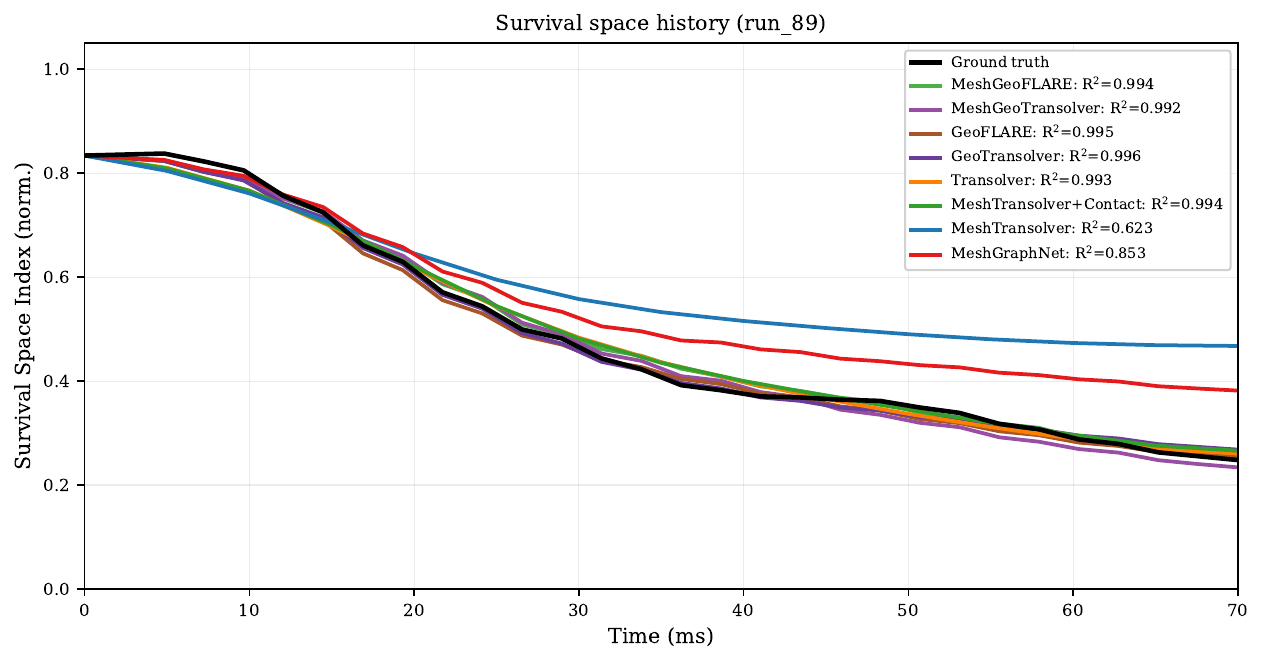}
\caption{Occupant survival-space distance over time for a representative test case. The strongest models track the temporal evolution of the downstream metric closely, including both compression and rebound. The survival distance axis is normalized.}
\label{fig:survival_space_curves}
\end{figure}

Figure~\ref{fig:survival_space_curves} shows predicted and ground-truth survival-distance histories for a representative test case. The strongest models reproduce not only the overall level of the signal, but also its temporal evolution through the compression and rebound phases. Table~\ref{tab:survival_space} summarises the survival-space error statistics over the full 25-sample test set.

\begin{table}[t]
\centering
\caption{Occupant survival-space error over the 25 in-distribution test samples. Mean error quantifies systematic bias, while standard deviation reflects consistency across the test set.}
\label{tab:survival_space}
\small
\begin{tabular}{lcc}
\toprule
Model & Mean error (mm) & Std error (mm) \\
\midrule
GeoTransolver & $\mathbf{-0.9}$ & $10.3$ \\
GeoFLARE & $-1.3$ & $10.6$ \\
Transolver & $-2.8$ & $11.5$ \\
\textit{MeshTransolver+Contact} & $-3.9$ & $11.7$ \\
\textit{MeshGeoFLARE} & $-4.2$ & $\mathbf{10.1}$ \\
\textit{MeshGeoTransolver} & $-4.3$ & $10.6$ \\
MeshGraphNet & $-5.6$ & $50.5$ \\
\textit{MeshTransolver} & $+46.0$ & $15.2$ \\
\bottomrule
\end{tabular}
\end{table}

The strongest hybrid models, \textit{MeshGeoFLARE} and \textit{MeshGeoTransolver}, achieve moderate negative bias ($-4.2$ and $-4.3$~mm, respectively) with the lowest standard deviation among all models ($\sim$10~mm), indicating strong consistency across the test set. Transolver also performs well on this downstream metric, which is consistent with its strong global field accuracy.

A particularly informative comparison is the one between \textit{MeshTransolver} and \textit{MeshTransolver+Contact}. Without explicit contact modelling, \textit{MeshTransolver} exhibits a large positive bias ($+46.0$~mm), meaning that it systematically predicts more residual occupant opening than the FE reference. This behavior is non-conservative from a safety perspective and is clearly inaccurate. Introducing the sparse contact block reduces the mean error to $-3.9$~mm and also lowers the spread, indicating that explicit contact modelling substantially improves the preservation of this downstream quantity.

MeshGraphNet shows a mean error close to zero on average, but a much larger standard deviation than the strongest hybrid models. This indicates that average agreement alone is not sufficient: consistency across design variants is equally important when the metric is used for design screening.

\begin{figure}[t]
\centering
\includegraphics[width=0.7\linewidth]{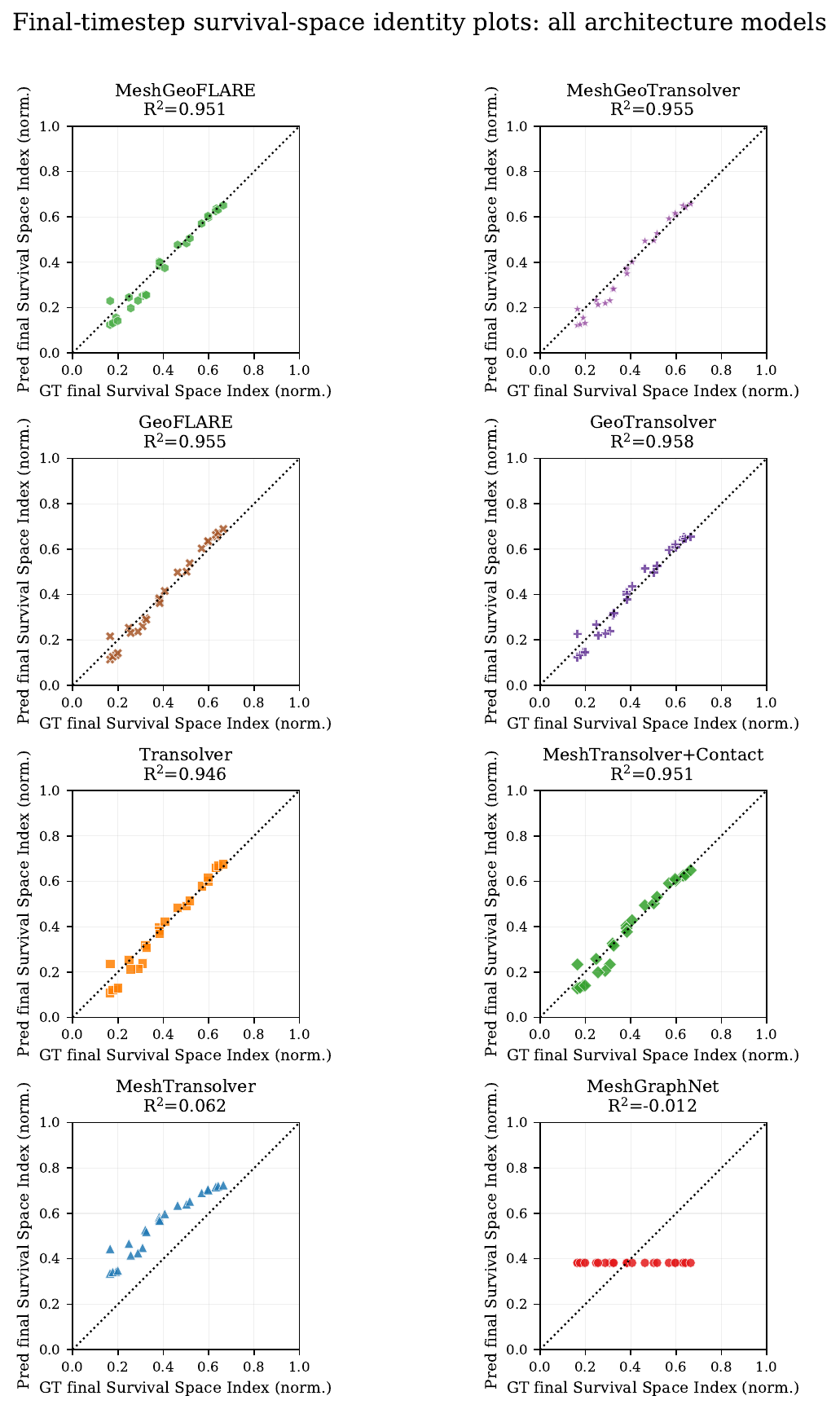}
\caption{Predicted versus reference final survival distance for the 25 in-distribution test samples. Stronger models remain closer to the diagonal, indicating lower bias and lower dispersion in the downstream safety metric. The survival distance axes are normalized.}
\label{fig:survival_scatter}
\end{figure}

Figure~\ref{fig:survival_scatter} further illustrates the agreement through predicted-versus-reference final survival distance. The strongest models cluster more closely around the diagonal, whereas weaker models exhibit larger dispersion away from the ideal agreement line. From an industrial perspective, this is important because both the magnitude and the sign of the survival-space error affect how intrusion is interpreted during DOE screening.

\subsection{Computational efficiency}
\label{sec:results_efficiency}

A key practical advantage of the proposed surrogate models is the drastic reduction in evaluation time. Once trained, the surrogate performs a full autoregressive rollout of the crash sequence in approximately 2.6 seconds per design on a single GPU.

For the reduced benchmark model considered in this work, a corresponding high-fidelity FE simulation using PAM-CRASH requires approximately 15 minutes on an HPC CPU setup. This already represents a speedup of approximately $350\times$.

More importantly, for full-scale industrial vehicle models, a single FE simulation typically requires on the order of 10–15 hours, depending on mesh resolution and solver configuration. In this regime, the surrogate achieves speedups of up to $2\times10^4$, reducing evaluation time from hours to seconds.

This difference in computational cost is what enables large-scale design-space exploration. While FE-based workflows are inherently limited in the number of designs that can be evaluated, the surrogate allows rapid screening of hundreds or thousands of design variants within practical time constraints.

\begin{table}[t]
\centering
\caption{Approximate computational cost per design. Surrogate runtime corresponds to a full autoregressive rollout.}
\label{tab:runtime}
\begin{tabular}{lcc}
\toprule
Method & Time per design & Hardware \\
\midrule
FE simulation (reduced model) & $\sim$15 min & HPC CPU cluster \\
FE simulation (full vehicle) & $\sim$10--15 h & HPC CPU cluster \\
Neural surrogate inference & 2.6 s & 1$\times$ GPU \\
\midrule
Speedup (reduced model) & $\sim$350$\times$ & -- \\
Speedup (full vehicle) & $\sim 10^{4}$--$2\times 10^{4}$ & -- \\
\bottomrule
\end{tabular}
\end{table}

The reported runtime corresponds to a full autoregressive rollout over the complete crash sequence for a single design (batch size = 1), including all predicted timesteps.

%% file: 07_discussion.tex
\section{Discussion}
\label{sec:discussion}

The benchmark supports three main empirical observations.
First, under the present benchmark and training protocol, hybrid mesh--attention architectures form the strongest-performing group.
Second, within the evaluated hybrid family, explicit sparse contact modelling produces the largest observed improvement.
Third, downstream safety-relevant evaluation provides information that is not fully captured by global nodal error alone.

\subsection{What hybrid mesh--attention architectures contribute}
\label{sec:disc_hybrid}

The ranking in Table~\ref{tab:id_results} and the temporal behaviour in Figure~\ref{fig:rmse_hybrid} indicate that local and global processing play complementary roles. Mesh-based propagation preserves neighbourhood-level structural detail, whereas the token-based global processor enables efficient long-range information exchange across the full vehicle. Under the evaluated configuration, the strongest-performing models are those that retain both mechanisms.

The geometry-aware hybrids, especially \textit{MeshGeoFLARE}, further suggest that geometric conditioning is most useful when embedded within a mesh-based backbone. In contrast to earlier iterations, the updated results indicate that geometry-aware attention alone can capture the dominant global crash kinematics under the shared hyperparameter configuration. \textit{GeoTransolver} and \textit{GeoFLARE} therefore achieve competitive scalar RMSE and survival-space bias. However, qualitative side-view comparisons (Figure~\ref{fig:side_view_field_regularization}) show that these models can introduce higher spatial noise and local deformation irregularities. This suggests that global tokenisation benefits from an explicit local mesh inductive bias when the objective is not only scalar accuracy, but also field regularity and structural interpretability for industrial crash analysis. For the present benchmark and configuration, the hybrid models provide the best balance of temporal accuracy, final-step precision, and physically interpretable displacement fields.

\subsection{Role and limits of contact modelling}
\label{sec:disc_contact}

The temporal ablation in Figure~\ref{fig:contact_ablation_temporal} shows that within the MeshTransolver comparison, explicit sparse contact modelling improves both overall rollout accuracy and, more importantly, the early-impact regime. The largest gain appears when contact first becomes dynamically important, which is consistent with the intended role of the generic sparse contact block: to introduce interaction information that is not encoded by the static mesh before the global processor propagates the disturbance through the structure.

What can be stated robustly is that the sparse contact block provides a useful and scalable representation of evolving interactions in the present full-vehicle setting. The results are also consistent with the architectural interpretation adopted in this work: pre-MPNN extracts a locally structured latent state on the mesh, the contact block injects contact-dependent corrections, the global processor propagates their effect across the structure, and the final post-MPNN stage refines the response on the stable structural mesh only.

At the same time, the present study does not claim to fully isolate all mechanisms behind the observed gain. The main conclusion is therefore empirical rather than mechanistic: sparse contact injection before global propagation is an effective design choice for autoregressive full-vehicle crash prediction under the current benchmark.

\subsection{Industrial interpretation of survival-space evaluation}
\label{sec:disc_survival}

Occupant survival space should be interpreted as a downstream industrial metric rather than as part of the training objective. This distinction matters because a model can exhibit acceptable average nodal error while still producing undesirable bias or variability in a safety-relevant quantity. In particular, with the present definition, positive survival-space errors indicate that the surrogate predicts a larger residual survival distance than the FE reference, corresponding to a non-conservative underestimation of intrusion. Negative errors indicate a smaller predicted survival distance and are therefore conservative.

The present results show that the strongest hybrid models not only achieve low global field error, but also maintain low bias and relatively low dispersion in survival-space error. This is an important result from an industrial perspective, because downstream design screening depends not only on average agreement with the FE reference, but also on whether the surrogate preserves the sign and consistency of the safety-relevant metric across design variants.

These observations also clarify why downstream evaluation is necessary in crash surrogates. Standard field-level metrics remain important, but they are not sufficient on their own when the intended use case is design assessment rather than visual field reconstruction only.

\subsection{Architectural capacity and computational cost}
\label{sec:disc_capacity}

The analysis of the evaluated architectures reveals a clear trade-off between model capacity, training cost, and final accuracy. As shown in Table~\ref{tab:model_capacity}, incorporating explicit geometry-aware global attention and sparse contact modelling increases the model capacity and the corresponding computational cost per epoch. The \textit{MeshGeoFLARE} and \textit{MeshGeoTransolver} variants, while more parameter-intensive, consistently achieve the lowest temporal evaluation error and best survival-space preservation. This indicates a representative accuracy--capacity trade-off within the architecture family, where a practitioner can choose between a more compact hybrid and a larger geometry- and contact-aware variant depending on production latency constraints and the required precision of the downstream evaluation criteria.

The contact block itself adds only a modest parameter overhead in the evaluated hybrids (for example, about 20K parameters in the base MeshTransolver comparison), which is small relative to the total model size. This is consistent with the intended design: the contact block acts as a lightweight correction mechanism rather than as a structurally dominant component.

In addition to parameter count and number of training epochs, the effective training cost is also influenced by the per-epoch computational cost of each architecture. As reported in Table~\ref{tab:model_capacity}, the wall-clock time per epoch varies significantly across model families.

Importantly, this cost does not scale directly with parameter count alone. Instead, it is strongly affected by the structure of the processor. In particular, token-based global attention introduces additional overhead due to slice--deslice operations and token interaction, while contact-aware hybrids incur extra cost from dynamic contact graph construction at each rollout step.

For example, although GeoFLARE and GeoTransolver have comparable parameter counts, their per-epoch training time differs noticeably, reflecting differences in the underlying attention mechanism. Similarly, hybrid models exhibit higher per-epoch cost than pure mesh-based models, but compensate for this through the improved field regularity and structural plausibility demonstrated in the qualitative analysis.

From a practical perspective, the relevant quantity is therefore the total time-to-solution. Under the common training protocol used here, the \textit{MeshTransolver} family achieves a favourable balance by combining moderate per-epoch cost with competitive accuracy, while larger geometry- and contact-aware variants trade higher computational cost for improved final precision and safety-metric preservation.

\subsection{Challenges and future work}
\label{sec:disc_limits}

Several challenges remain. The current study focuses on a single load case, namely full-vehicle lateral pole impact.

\begin{figure}[t]
  \centering
  \includegraphics[width=\linewidth]{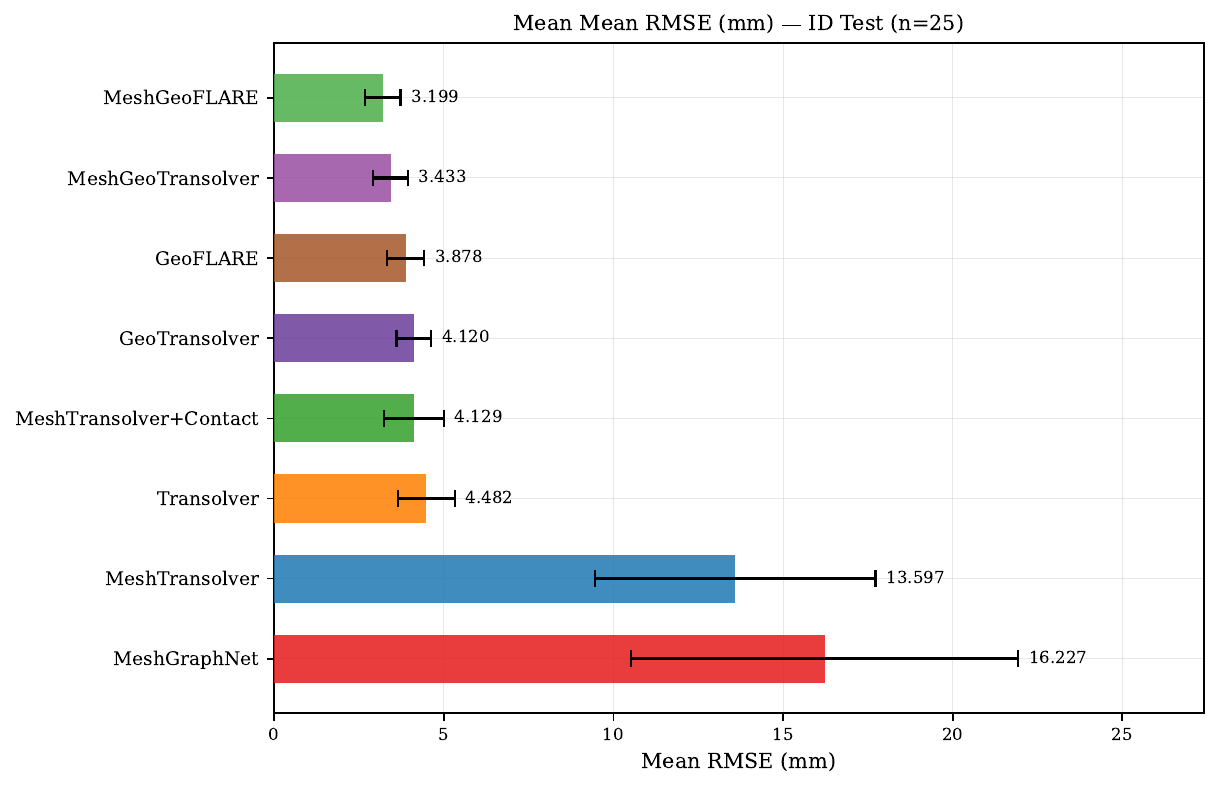}
  \caption{Summary ranking of the evaluated models on the full-vehicle benchmark, ordered by $\mathrm{RMSE}_{\mu}$. Hybrid mesh--attention architectures occupy the top positions, combining strong quantitative performance with the field regularity shown in Figure~\ref{fig:side_view_field_regularization}.}
  \label{fig:model_ranking}
\end{figure}

In addition, the present evaluation is restricted to an in-distribution design-space-aware split, so broader out-of-distribution assessment remains an important next step. Figure~\ref{fig:model_ranking} summarises the overall ranking and reinforces the same conclusion drawn from the main results table: hybrid mesh--attention architectures form the strongest group, while geometry-only attention models remain competitive in terms of scalar metrics but are qualitatively distinct in terms of field regularity.

Future work should therefore address additional crash scenarios, stronger out-of-distribution evaluation, safety-aware training objectives, uncertainty quantification, and further refinement of contact modelling under more severe distribution shift.

A further design choice in this study is to prioritise a controlled common training protocol over exhaustive per-model hyperparameter optimization. This improves comparability across architecture families and avoids confounding architectural effects with unequal tuning budgets. Therefore, the reported results should be interpreted as a fixed-protocol architectural comparison rather than as an exhaustive per-model optimized leaderboard.

%% file: 08_conclusion.tex
\section{Conclusion}
\label{sec:conclusion}

This work presented and evaluated fully autoregressive neural surrogates for full-vehicle lateral pole-impact simulation. On the present benchmark, the strongest-performing models are hybrid mesh--attention architectures that combine local mesh-based propagation with a global token-based processor, with the best results obtained by the geometry-aware hybrid variant MeshGeoFLARE.

Within the evaluated architecture family, explicit sparse contact injection substantially improves autoregressive prediction, particularly in the early-impact regime. Beyond global field errors, the study also shows that occupant survival space provides a meaningful downstream safety-relevant metric for assessing practical utility in crash-engineering workflows.

Taken together, the results suggest that combining mesh-local structure, global latent communication, and contact-aware correction is a promising design direction for industrial crash surrogates. Under the shared hyperparameter configuration used in this study, geometry-aware attention baselines achieve competitive quantitative accuracy, while hybrid mesh--attention models remain preferable for industrial deployment as they better preserve field regularity and physically interpretable deformation patterns. At the same time, these conclusions are established on a single-load-case, in-distribution DOE-aware benchmark; broader out-of-distribution evaluation and model-specific hyperparameter tuning remain important directions for future work.

Overall, the proposed framework advances crash-surrogate modelling toward deployment-oriented design-space screening by combining fully autoregressive rollout, scalable hybrid architectures, and downstream safety-relevant evaluation.